\documentclass[12pt]{amsart}
\usepackage[foot]{amsaddr}

\usepackage{cite}
\usepackage{lineno}
\usepackage{amsmath,amsthm,amssymb}
\usepackage{enumerate}
\usepackage{color}
\usepackage{graphicx}
\usepackage{kotex}
\usepackage{caption}
\usepackage[labelformat=simple]{subcaption}
\usepackage[left=25.4mm,right=25.4mm,top=30.0mm,bottom=25.4mm,a4paper]{geometry}
\newcommand{\gen}{\mathrm{gen}}

\def\alphaBar{\langle \alpha \rangle_T}
\def\rhoBar{\langle \rho \rangle_x}
\def\kappaBar{\langle \kappa \rangle_x}

\def\alphaBar{\overline \alpha }
\def\rhoBar{\overline \rho}
\def\kappaBar{\overline \kappa}
\def\Qh{Q_h}
\def\Qc{Q_c}
\def\QhBar{\overline \Qh }
\def\QcBar{\overline \Qc }

\title{Thermoelectric degrees of freedom determining thermoelectric efficiency}
\author{Byungki Ryu$^{\dagger,*}$}
\author{Jaywan Chung$^{\dagger,*}$}
\author{SuDong Park}
\address[B Ryu, J Chung, \and SD Park  ]{Energy Conversion Research Center, Korea Electrotechnology Research Institute (KERI), 12, Bulmosanro 10beon-gil, Seongsan-gu, Changwon, 51543, Republic of Korea}
\address[$^{\dagger}$]{B.R. and J.C. contributed equally to this work}
\email[B. Ryu (Corresponding Author)]{byungkiryu@keri.re.kr, jchung@keri.re.kr}
\date{\today}

\begin{document}

%\setpagewiselinenumbers
%\linenumbers

%\onecolumn
%\twocolumn

\maketitle

\begin{abstract}
Thermal energy can be directly converted to electrical energy as a result of thermoelectric effects.
Because this conversion realises clean energy technology, such as waste heat recovery and energy harvesting, substantial efforts have been made to search for thermoelectric materials.
Under the belief that the material figure of merit $zT$ represents the energy conversion efficiencies of thermoelectric devices, various high peak-$zT$ materials have been explored for half a century.
However, thermoelectric properties vary greatly with temperature $T$, so the single value $zT$ does not represent device efficiency accurately.
%\textcolor{red}{(1) Here we show that efficiency of thermoelectric conversion has three degrees of freedom and efficiency is completely determined by three parameters $Z_\gen$, $\tau$, and $\beta$.} 
Here we show that the efficiency of thermoelectric conversion is completely determined by \emph{three} parameters $Z_\gen$, $\tau$, and $\beta$, which we call the \emph{thermoelectric degrees of freedom}.
%\textcolor{green}{and there are three degrees of freedom.}
The $Z_\gen$, which is an average of material properties, is a generalisation of the traditional figure of merit.
The $\tau$ and $\beta$, which reflect the gradients of the material properties, are proportional to escaped heat caused by the Thomson effect and asymmetric Joule heat, respectively.
Our finding proposes new directions for achieving high thermoelectric efficiency; increasing one of the thermoelectric degrees of freedom results in higher efficiency. 
For example, thermoelectric efficiency can be enhanced up to 176\% by tuning the thermoelectric degrees of freedom in segmented legs, compared to the best efficiency of single-material legs.
\end{abstract}

%\twocolumn

%\subsection*{Introduction}
Thermoelectric device can convert heat to electricity by generating electrical voltage across legs operating between a hot side, with temperature $T_h$, and cold side, with temperature $T_c$.
In general, the nonlocality of the thermoelectric equation allows no analytical expression for energy conversion efficiency \cite{goupil2015continuum, chung2014nonlocal}.
%However, \emph{if} the thermoelectric material properties \emph{do not vary with temperature $T$},
However, \emph{if} thermoelectric properites of materials \emph{do not vary with temperature $T$},
the maximum efficiency of thermoelectric conversion $\eta_{\rm max}$ is determined by the dimensionless figure of merit ${ z} T_m =  \frac{\alpha^2}{\rho \kappa}  T_m$, 
where $\alpha$ is the Seebeck coefficient, $\rho$ is the electrical resistivity, $\kappa$ is the thermal conductivity, and $T_m$ is the mean of $T_h$ and $T_c$.
%Preciesly 
The $\eta_{\rm \,max}$ for temperature-independent (or constant) material properties is given by the well-known classical formula $\eta^{\rm const}_{\rm max} = \frac{\Delta T}{T_h}\frac{\sqrt{1+zT_m}-1}{\sqrt{1+zT_m}+ T_c / T_h}$ where $\Delta T =T_h-T_c$.
In this classical formula, higher $zT_m$ implies higher efficiency. With this observation, extensive research on materials has been conducted, improving peak $zT$ from below 1 to above 2.6 in half a century \cite{zhao2014ultralow}.
Low thermal conductivity and high power factor ($\alpha^2 / \rho$) have been achieved by using anharmonic phonon structure, defective structures from atomic scale to micro scale, non-parabolic band structure, electronic structure distortion, and Fermi-level tuning 
\cite{zhao2014ultralow, biswas2012high,kim2015dense, snyder2008complex,heremans2008enhancement,pei2011convergence,liu2012convergence}.

\begin{figure}[ht]
\centering \includegraphics[width=0.5\textwidth]{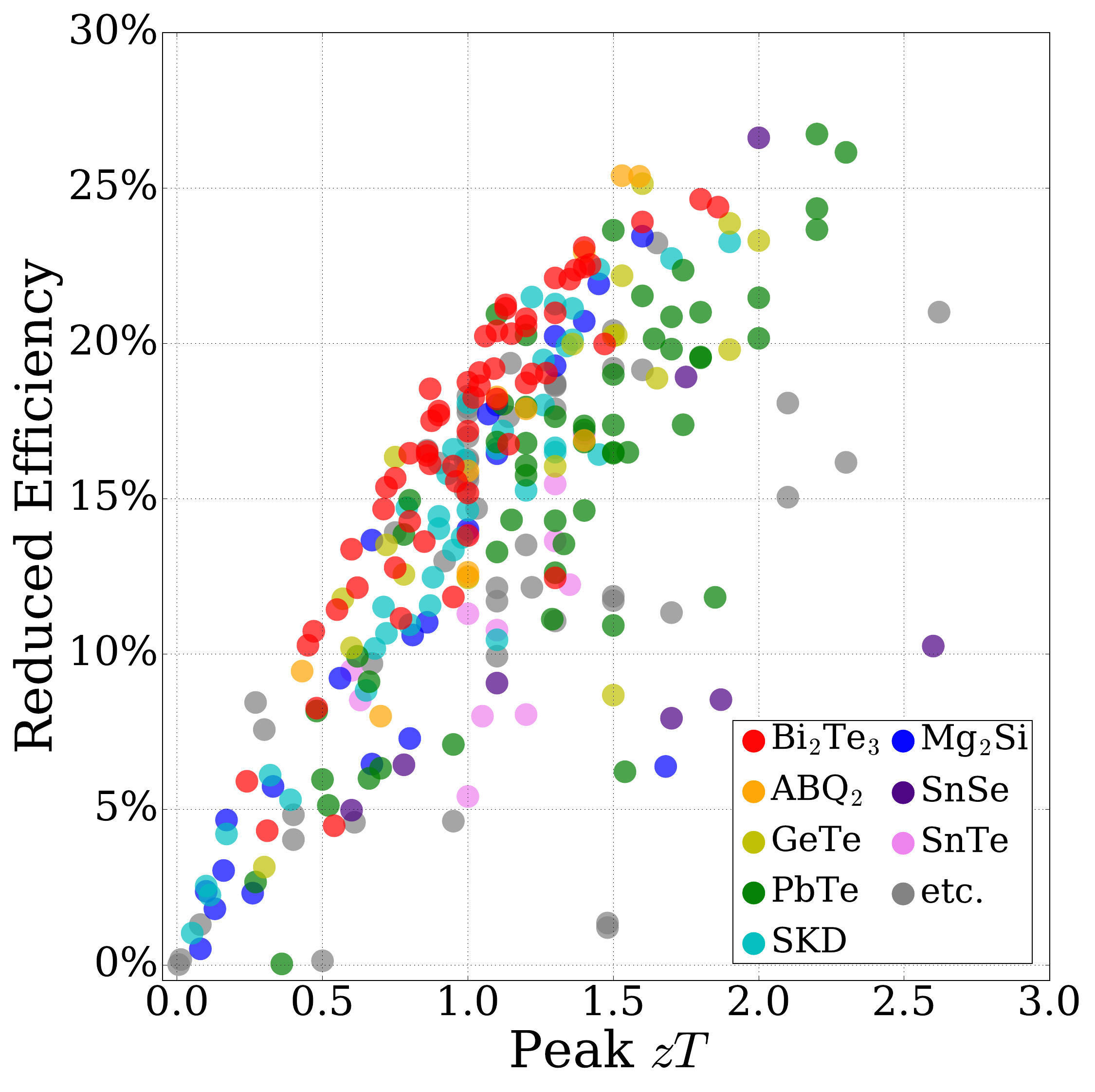}
\caption{Relationship between the reduced efficiency $\eta_{\rm red}=\eta_{\rm max}/\eta_{\rm Carnot}$ and the peak $zT$ for 276 reported materials, 
including Bi$_2$Te$_3$, PbTe, SnSe, and so on.
%including Bi$_2$Te$_3$, ABQ$_2$, GeTe, PbTe, skutterudite (SKD), Mg$_2$Si, SnSe, SnTe-based materials, and so on. 
%\textcolor{red}{:for material information, see \S1 in SI.}
}\label{literatureAnalysis}
\end{figure}

However, material properties \emph{do} vary greatly with temperature $T$, hence a higher peak $zT$ no longer implies higher device efficiency.
In Figure \ref{literatureAnalysis}, the relationship between the peak $zT$ and the numerical maximum efficiency divided by the Carnot efficiency $\eta_{\rm Carnot} = \frac{\Delta T}{T_h}$ is shown for 276 reported materials under available temperature ranges; see \S1 and \S2 in Supplementary Information(SI) for details. The $\eta_{\rm red}$ can vary by a factor of five between materials even if the peak $zT$ is the same. 
The strong correlation among similar materials in Figure \ref{literatureAnalysis} is misleading; 
the correlation exists not because
the peak $zT$ is predictive, but because the $zT$ ignores the temperature dependencies of the similar materials altogether.

To overcome the limitation of the peak $zT$, various average parameters have been proposed.
Ioffe and Borrero \cite{ioffe1957semiconductor,borrego1961optimum, sherman1960calculation} suggested an average figure of merit $z_{\mathrm{av}} = \frac{\left<\alpha\right>^2}{\left<\rho\kappa\right>}$, where the bracket $\left<\cdot\right>$ indicates averaging over $T$, from \emph{constant heat-current approximation}.
However, the average $z$ or $Z$ does not predict efficiency accurately, due to a disregard for the Thomson effect and the non-uniformity of Joule heat, as well as poor approximation of the temperature distribution \cite{sherman1960calculation, sunderland1964influence, wee2011analysis, sandoz2013thomson, ju2017revisiting}. 
Snyder and Snyder \cite{snyder2017figure} defined \emph{device $ZT$} in terms of $\eta_{\rm max}$: $(ZT)_{\rm dev} := \left( \frac{T_h-T_c(1-\eta_{\rm max})}{T_h(1-\eta_{\rm max}) - T_c}  \right)^2 -1$. However, while the $(ZT)_{\rm dev}$ satisfies the classical efficiency formula, it gives no method to compute $\eta_{\rm max}$.
Kim et al.\cite{kim2015relationship, kim2015efficiency, kim2017bridge} suggested an efficiency formula using \emph{engineering $ZT$} and Thomson correction parameters. Their efficiency formula accounts for strong temperature dependence of material properties; for example, SnSe has the peak $zT$ value of 2.6 at 923 K while the $zT$ vanishes near the room temperature \cite{zhao2014ultralow}. However, it still contains about 9\% error for SnSe due to the \emph{linear approximation of temperature}.
One may consider the compatibility factor $s = \frac{ \sqrt{1+zT}-1}{\alpha T}$, suggested by Snyder and Ursell \cite{snyder2003thermoelectric}. As a function of $T$, the uniformity of $s(T)$ may indicate high efficiency. However, the $s$ is not a figure of merit; higher $s$ does not imply higher efficiency, while higher $zT$ partially does. In summary, no \emph{single} value has been successful in predicting thermoelectric conversion efficiency as a figure of merit.

\subsection*{Main Result}
\begin{figure}[ht]
\centering \includegraphics[width=0.8\textwidth]{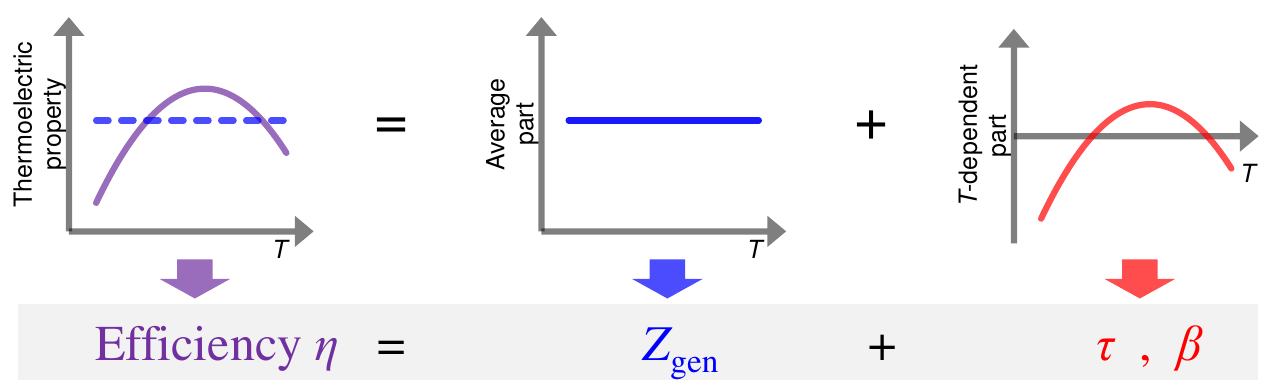}
\caption{A schematic diagram explaining the relationship between temperature-dependent material properties and thermoelectric degrees of freedom $Z_\gen$, $\tau$, and $\beta$.
The $Z_\gen$ represents the averaged part of material properties, while the $\tau$ and $\beta$ represent the gradient parts of them.
}\label{fig-schematic}
\end{figure}

\begin{figure}[ht]
\centering \includegraphics[width=0.5\textwidth]{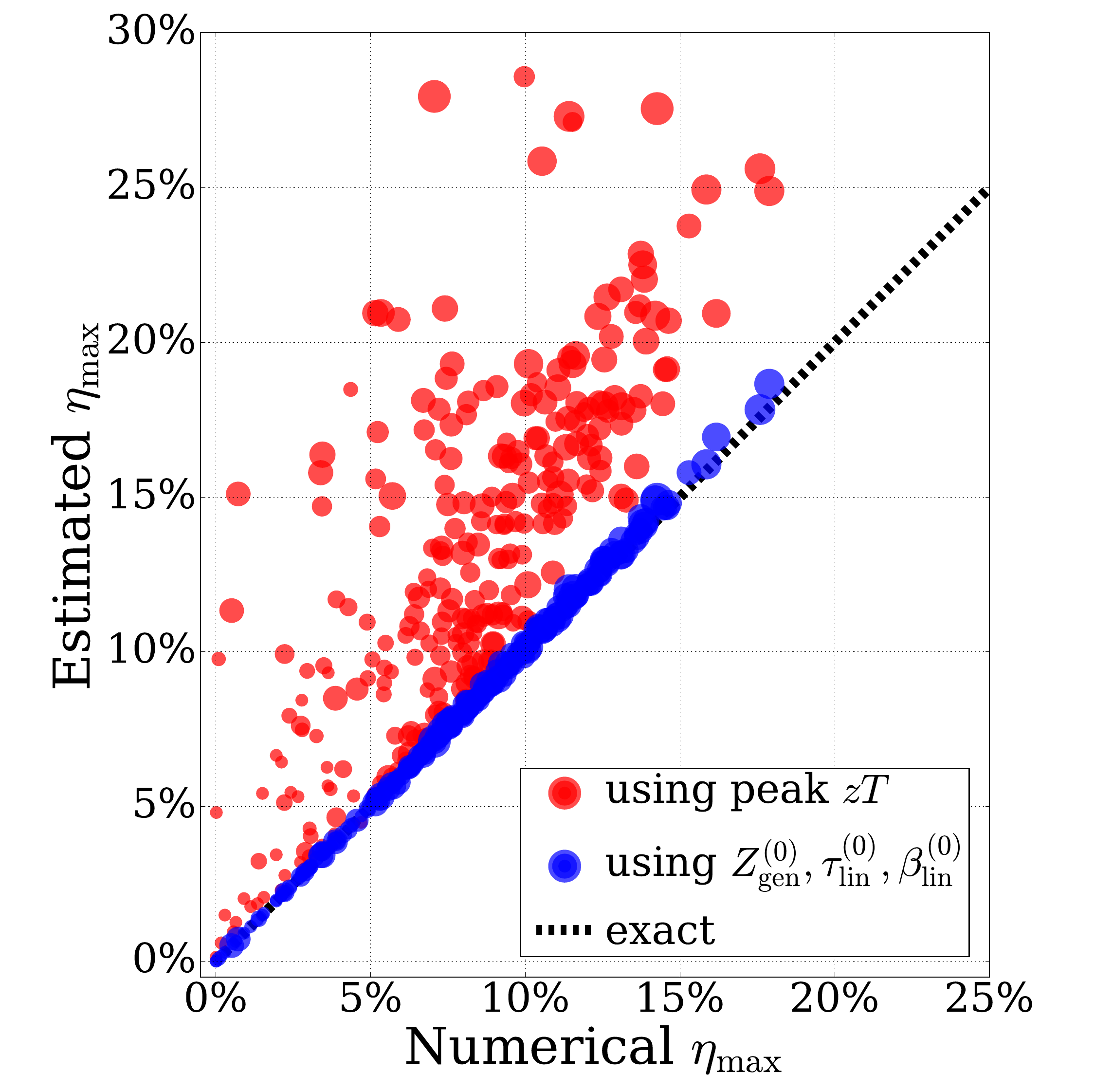}
\caption{Comparison of two estimation methods for maximum efficiency; one uses the classical formula $\eta_{\rm max}^{\rm const}$ with peak $zT$ (red circle), and the other uses our efficiency formula $\eta_{\rm max}^{\rm gen}$ in equation \eqref{ETAnew}, with the \emph{one-shot approximation} of the thermoelectric degrees of freedom in equation \eqref{one-shot-approx} (blue circle). The estimated values are compared with the exact, numerical maximum efficiency for 276 materials under available temperature range. The marker size is proportional to the peak $zT$.
}\label{fig-estimation}
\end{figure}

\begin{figure}[ht]
\centering
\begin{subfigure}{0.45\textwidth}
\caption{}
\includegraphics[width=\textwidth]{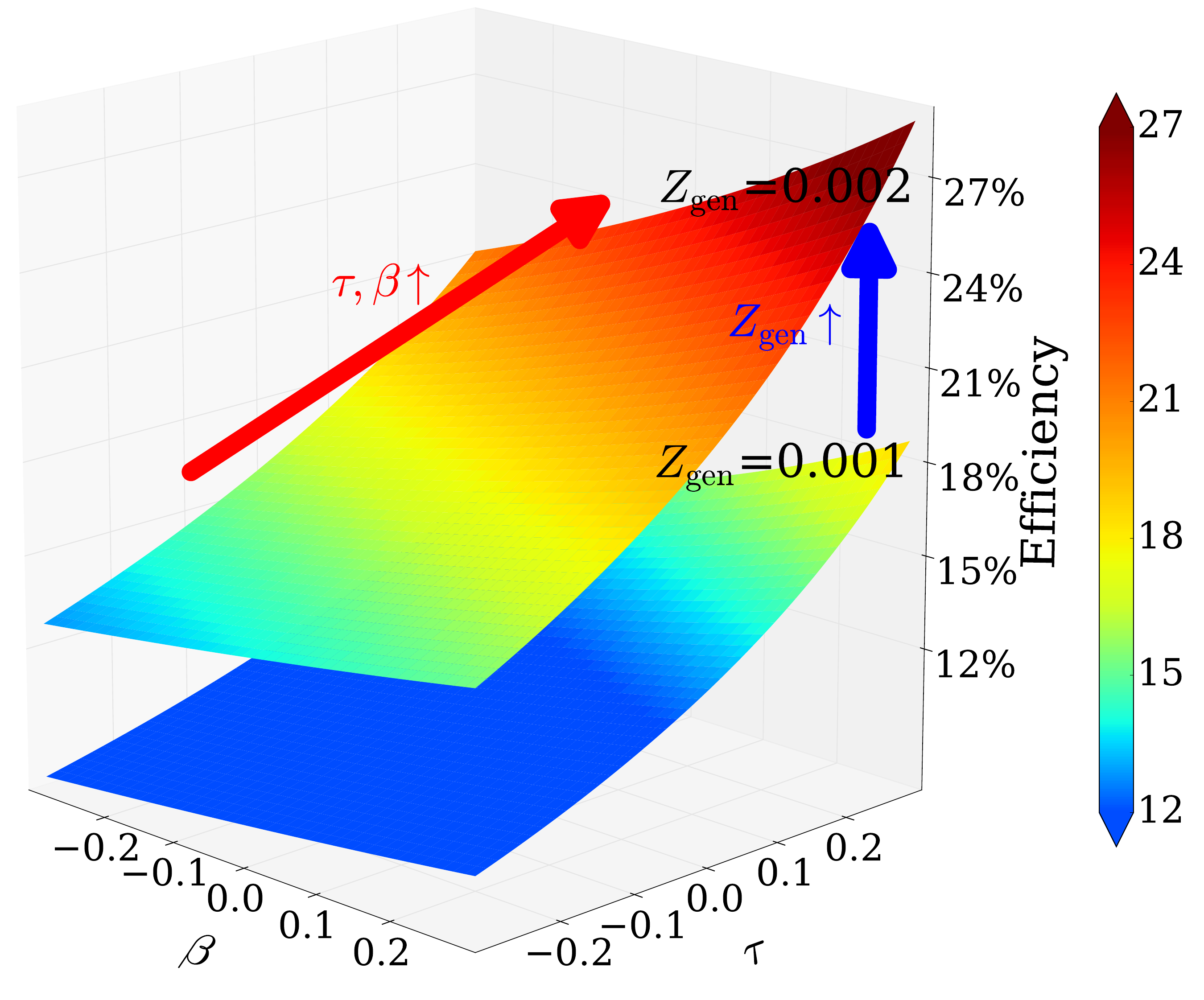}
\end{subfigure} \\
\begin{subfigure}[b]{0.45\textwidth}
\caption{}
\includegraphics[width=\textwidth]{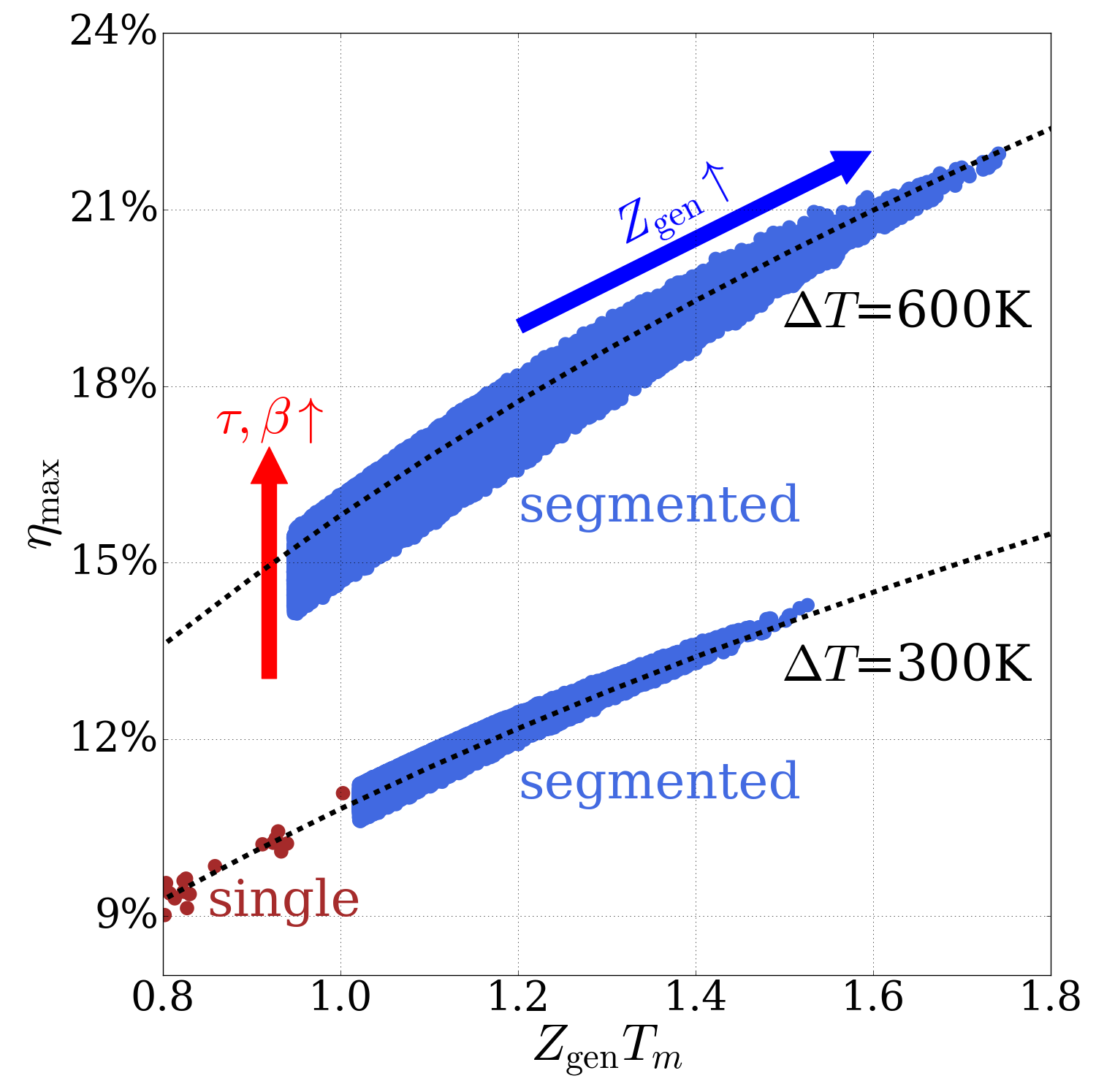}
\end{subfigure} \hfill
\begin{subfigure}[b]{0.45\textwidth}
\caption{}
\includegraphics[width=\textwidth]{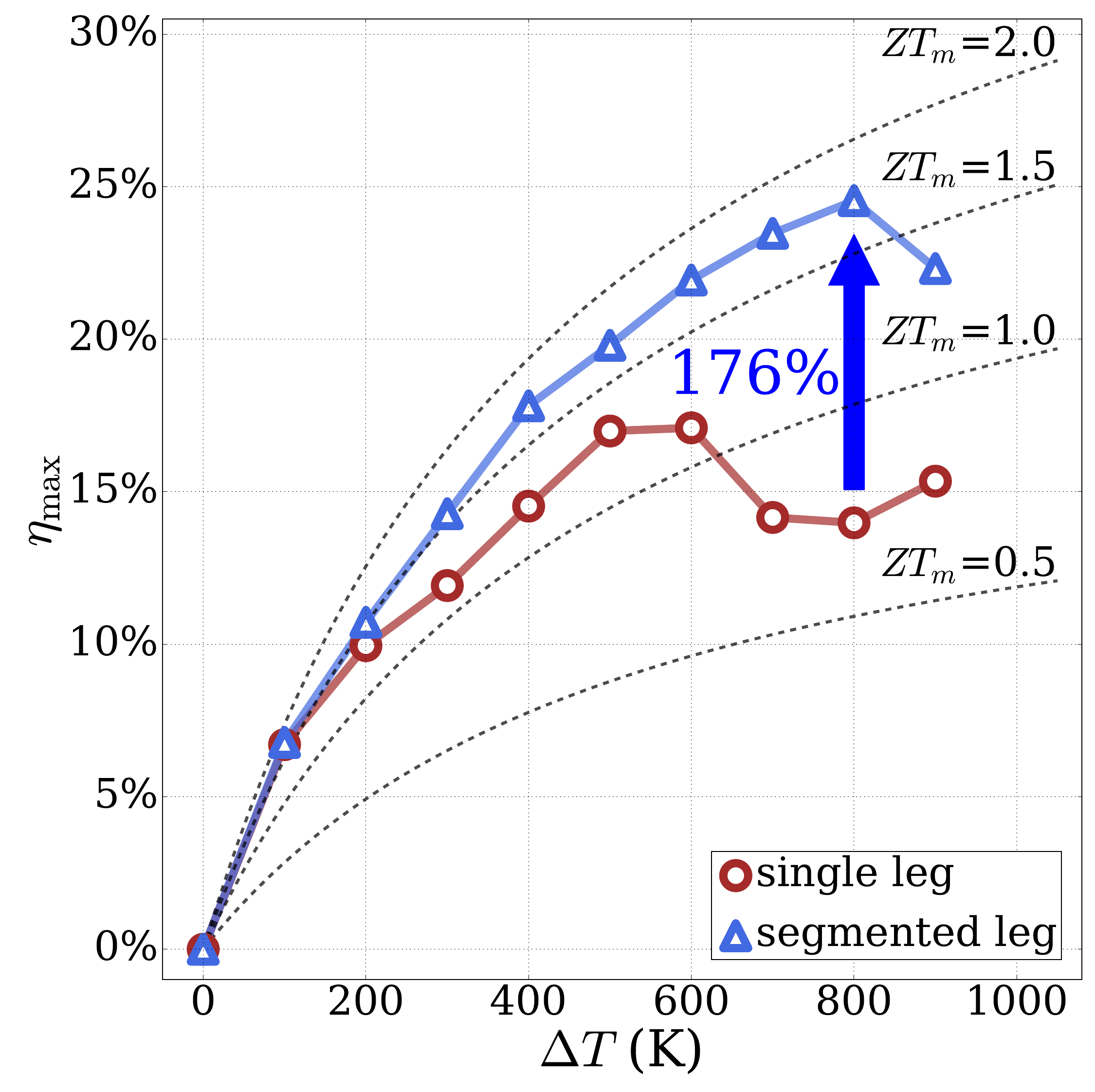}
\end{subfigure} \hfill
\caption{
(a) Efficiency surface $\eta_{\rm max}^{\rm gen} ( Z_\gen, \beta, \tau )$ in equation \eqref{ETAnew} is drawn for $Z_\gen = 0.002 K^{-1}$ and $0.001 K^{-1}$, for fixed $T_h = 900 K$ and $T_c = 300 K$. Improving one of three thermoelectric degrees of freedom $Z_{\rm gen},\tau,\beta$ increases the efficiency.
(b) Numerical maximum efficiencies for single-material legs (brown circle) and segmented legs (blue circle) for $T_c=300K$ and $\Delta T = 300 K$ or $600 K$. For the segmented legs, we consider $18^5 =1,889,568$ configurations up to 5-stage segmentation consisting of 18 candidate materials; for material information, see \S1 in SI. Here only the top 100,000 configurations are shown. Notice that the segmentation can control the thermoelectric degrees of freedom.
(c) Maximum possible efficiencies among single-material legs (brown open circle) and segmented legs (blue open triangle) for $T_c=300 K$ and given $\Delta T$. The leg segmentation can enhance the maximum efficiency up to 176\%.
}\label{fig-Ztaubeta-manage}
\end{figure}

In this paper, we show that thermoelectric conversion efficiency $\eta$ is \emph{completely determined} by \emph{three} independent parameters $Z_{\gen}$, $\tau$, and $\beta$. 
Because they determine the performance of thermoelectric devices, we call them the \emph{thermoelectric degrees of freedom}.
The $Z_\gen$ is an average of material properties, and it generalises the traditional figure of merit.
The additional degrees of freedom $\tau$ and $\beta$, which reflect the gradients of the material properties, are proportional to the escaped heat caused by the Thomson effect and the asymmetric Joule heat respectively.
We omit exact involved definitions of $Z_{\gen}$, $\tau$, and $\beta$ here because there is a simple but accurate approximation formula in equation \eqref{one-shot-approx} for them; for the exact definitions, see \S5 and \S8 in SI.
The schematic diagram in Figure \ref{fig-schematic} shows the relationship between material properties and the thermoelectric degrees of freedom, where the temperature-dependent material properties are decomposed into averaged and gradient parts, and they are represented by the degrees of freedom.
Efficiency is a function of the thermoelectric degrees of freedom when the electric current $I$ is given: $\eta = \eta(Z_\gen, \tau, \beta |I)$. Furthermore, each degree of freedom is a figure of merit because $\eta(Z_\gen, \tau, \beta)$ is monotone increasing in each variable; see \S9 in SI. 
The maximum efficiency can be simply approximated by the following formula:
\begin{equation}\label{ETAnew}
\eta_{\rm max}  = \eta_{\rm max} ( Z_\gen, \tau, \beta | I =  I_{\rm opt})  \approx  \eta_{\rm max}^{\rm gen} (Z_\gen, \tau, \beta) :=\frac{\Delta T}{T_h'} \frac{\sqrt{1+Z_\gen T_m'}-1}{\sqrt{1+Z_\gen T_m'}+ \frac{T_c'}{T_h'}},
\end{equation}
where $I_{\rm opt}$ is the optimal current giving the maximum efficiency, $T_h' = T_h - \tau \Delta T$, $T_c' = T_c - \left( \tau + \beta \right) \Delta T$, and $T_m' = (T_h' + T_c')/2$.
The general formula $\eta_{\rm max}^{\rm gen}$ is identical to the classical formula $\eta_{\rm max}^{\rm const}$ except for the modification of temperature parameters.
While exact computation of the $Z_{\gen}$, $\tau$, and $\beta$ requires temperature distribution inside the device,
they can be easily estimated using \emph{one-shot approximation} assuming constant heat current and linear thermoelectric properties; see \S10 in SI. The one-shot approximation gives
\begin{equation}\label{one-shot-approx}
Z_\gen \approx Z_\gen^{(0)} := \frac{ \left( \int_{T_c}^{T_h} \alpha \,dT \right)^2 }{\Delta T \,\int_{T_c}^{T_h} \rho \kappa \,dT}, \quad \tau \approx \tau^{(0)}_{\rm lin} := -\frac{1}{3} \frac{\alpha_h - \alpha_c}{\alpha_h + \alpha_c}, \quad \beta \approx \beta^{(0)}_{\rm lin} := \frac{1}{3}\frac{ \left( \rho\kappa \right)_h -\left( \rho\kappa \right)_c}{ \left( \rho\kappa \right)_h +\left( \rho\kappa \right)_c },
\end{equation}
where the subscripts $h$ and $c$ indicate the material properties evaluated at $T_h$ and $T_c$ respectively. When the material properties are temperature-independent, the above formulas, $\eta_{\rm max}^{\rm gen}$, $Z_\gen^{(0)}$, $\tau^{(0)}_{\rm lin}$ and $\beta^{(0)}_{\rm lin}$ in equation \eqref{ETAnew} and \eqref{one-shot-approx}, become $\eta_{\rm max}^{\rm const}$ and the traditional figure of merit, $Z_\gen=z$, with vanishing $\tau$ and $\beta$.
The simple formula in equation \eqref{ETAnew} with equation \eqref{one-shot-approx} predicts maximum efficiency with high accuracy; see Figure \ref{fig-estimation}.
Besides predicting efficiency, the thermoelectric degrees of freedom suggest new directions for improving it. Since each of $Z_\gen$, $\tau$, and $\beta$ is a figure of merit, by increasing one of them, we can improve efficiency; see Figure \ref{fig-Ztaubeta-manage}(a),(b). Segmenting a leg with different materials is a way of tuning the degrees of freedom and increasing efficiency. In this way, efficiency can be enhanced up to 170\% compared to single-material legs; see Figure \ref{fig-Ztaubeta-manage}(c).

In the following sections, we sketch derivation of the thermoelectric degrees of freedom, and give more practical applications of them.
Full derivation can be found in SI.

%\subsection*{Main Ideas for Derivation}

\subsection*{Temperature Distribution}
The thermoelectric effect is expressed in terms of electric current density $J$ and heat current density $J^Q$: $J = \sigma ( E - \alpha \nabla T)$ and $J^Q =  \alpha T J - \kappa \nabla T$ where $E$ is the electric field. Applying the charge and energy conservation laws to $J$ and $J^Q$,
%and assuming \emph{one-dimensional} circuit case,
we can obtain the thermoelectric differential equation of temperature $T$ in a one-dimensional thermoelectric leg \cite{chung2014nonlocal, goupil2015continuum}:
\begin{equation}\label{TEQ-PDE-1D}
\frac{d}{dx} \left(\kappa \frac{dT}{dx} \right) + \rho J^2 -T \frac{d \alpha}{dT} \frac{dT}{dx} J = 0
\end{equation}
where $x$ is the spatial coordinate inside the leg. The left-hand side of equation \eqref{TEQ-PDE-1D} is composed of thermal diffusion, Joule heat generation, and Thomson heat generation. 
Here, we obtain an \emph{integral} equation for $T=T(x)$ by integrating the equation \eqref{TEQ-PDE-1D} twice. The integral equation is of the form $T = \varphi [T]$ where $\varphi$ is an integral operator; see \S 7 in SI. With this relation, the exact $T$ can be obtained via fixed-point iteration $T_{n+1} = \varphi [T_{n}] $ \cite{burden2010numerical}.
An approximate $T$ can be obtained from $T = \varphi [T^{(0)}]$, where $T^{(0)}$ is the solution of equation \eqref{TEQ-PDE-1D} when $J=0$.
Once the temperature distribution $T$ is found, the thermoelectric performance and efficiency can be easily computed.

\subsection*{Electrical Power}
To compute open-circuit voltage $V$, electrical resistance $R$, and thermal resistance $1/K$, 
it is natural to integrate the material properties on the spatial coordinate $x$ inside the leg (not on $T$) because the electric current and the heat current flow through the leg. Hence we define average parameters of material properties as 
$\alphaBar := \frac{ V }{\Delta T} = \int \alpha \frac{dT}{dx} \,dx$,
$\rhoBar := \frac{A}{L} R = \frac{1}{L} \int \rho \,dx$,
and $\frac{1}{\kappaBar} := \frac{A}{L} \frac{1}{K} = \frac{1}{L} \int \frac{1}{\kappa} \,dx$, 
where $L$ and $A$ are the length and area of the leg. 
The electric current is determined as $I = \frac{V}{ R \left( 1+ \gamma \right) }$, where $\gamma$ is the ratio of the load resistance $R_{\rm L}$ outside the device to the resistance of the thermoelectric leg $R$. The power delivered to the load is $P = I^2 R_{\rm L} = I ( V - I R ) = \frac{ \alphaBar^2 }{\rhoBar} \frac{\Delta T^2 }{L/A} \frac{\gamma}{\left( 1+\gamma \right)^2 }$, which is maximised near $\gamma = 1 $. Using the average properties, we define the \emph{general device power factor} as
${P\!F}_{\rm \!gen} := \frac{\alphaBar^2}{\rhoBar}$ and the \emph{general device figure of merit} as $Z_\gen = \frac{\alpha^2}{RK} = \frac{\alphaBar^2}{\rhoBar \,\kappaBar}$.

\subsection*{Heat Current}
If the material properties do not depend on $T$, the heat current of the hot and cold sides are determined by the average parameters:
$\QhBar =  K \Delta T + I \alphaBar T_h  - \frac{1}{2} I^2 R $ and
$\QcBar =  K \Delta T + I \alphaBar T_c  + \frac{1}{2} I^2 R $.
The power delivered outside is $P =\QhBar - \QcBar $. If the material properties \emph{depend on $T$}, the heat currents change to other values $Q_h$ and $Q_c$ but their difference remains unchanged, indicating that the power remains unchanged as well: $P = Q_h - Q_c = \QhBar - \QcBar$.
Thus, we have a relation $P = \Qh - \Qc = \left( \QhBar -B \right) - \left( \QcBar - B \right)$, implying that both of the heat currents are shifted by the same \emph{backward heat current $B$}.
We prove that $B$ is determined from the gradient parameters $\tau$ and $\beta$:
$B = \left(I \alphaBar  \Delta T \right) \tau   + \left( \frac{1}{2} I^2 R \right) \beta$; see \S8 in SI. Hence the heat currents at the boundaries are determined as
\begin{equation}\label{QhQc}
\begin{aligned}
Q_h &= K \Delta T + I \alphaBar ( T_h - \tau \Delta T )    - \frac{1}{2} I^2 R ( 1+\beta ), \\
Q_c &= K \Delta T + I \alphaBar ( T_c - \tau \Delta T )  + \frac{1}{2} I^2 R ( 1-\beta ).
\end{aligned}
\end{equation}
Note that the heat currents are changed by the effective Thomson heat flow $(-I \alphaBar \Delta T )\tau$ and the asymmetric Joule heat escape $(-\frac{1}{2} I^2 R) \beta$ due to the temperature dependency.

\subsection*{Efficiency}
Thermoelectric efficiency is defined as $\eta := \frac{Q_h - Q_c }{Q_h}$. From equation \eqref{QhQc}, we can verify that the dimensionless heat current, $\frac{Q_h}{K\Delta T}$ and $\frac{Q_c}{K\Delta T}$, and the efficiency are determined by five parameters $Z_\gen$, $\tau$, $\beta$, $\Delta T$, and $\gamma$. Furthermore, for fixed $\Delta T$ and $\gamma$, the efficiency is monotone increasing with respect to $Z_\gen$, $\tau$, and $\beta$ (see Figure \ref{fig-Ztaubeta-manage}(a)), which implies that each of the degrees of freedom $Z_\gen$, $\tau$, and $\beta$ is a figure of merit.

Although the degrees of freedom depend on $\gamma$, the dependency is negligible near the $\gamma_{\rm max}$ where the maximum efficiency occurs, because the temperature distribution hardly changes for $\gamma$ in most thermoelectric materials. Hence we may assume that the degrees of freedom in $\eta(Z_{\rm gen},\tau,\beta | \gamma)$ are fixed, independent values and maximise the $\eta$ only for $\gamma$ to find an approximate maximum efficiency. In this way we have the simple approximate formula $\eta_{\rm max}^{\rm gen}$ in equation \eqref{ETAnew} when $\gamma$ is near the $\gamma_{\rm max}^{\rm gen} :=  \sqrt{1 + Z_\gen T_m' }$; see \S9 in SI.
 
\subsection*{One-Shot Approximation}
The thermoelectric degrees of freedom can be directly estimated from the material properties using formula equation \eqref{one-shot-approx} for  $Z_\gen^{(0)}$, $\tau^{(0)}_{\rm lin}$ and $\beta^{(0)}_{\rm lin}$.
To derive the formula, we assume the temperature distribution inside the leg to be $T^{(0)}$, which is the temperature distribution for the $J=0$ case (i.e., $- \kappa \frac{dT}{dx} = \text{const.} $); the superscript ${(0)}$ means we use the $J=0$ case. Furthermore, we assume the material properties $\alpha$ and $\rho \times \kappa$ are linear with respect to $T$; the subscripts in $\tau^{(0)}_{\rm lin}$ and $\beta^{(0)}_{\rm lin}$ emphasize this linearity.
We then have the formula equation \eqref{one-shot-approx}; see \S10 and \S11 in SI for its derivation and accuracy. The general device power factor can be similarly estimated as 
$ {P\!F}_{\rm \!gen}^{(0)} =Z_\gen^{(0)} \,\int \kappa dT$.
This \emph{one-shot approximation} clarifies the relationship between the material properties and the thermoelectric efficiency; for example, because $\tau \approx -\frac{1}{3} \frac{\alpha_h - \alpha_c}{\alpha_h + \alpha_c}$, the $\tau$ (and the efficiency) can be enhanced if the $\alpha(T)$ declines more rapidly on $T$.
The one-shot approximation is accurate even for segmented devices. The performance of two-stage segmented device composed of SnSe and BiSbTe at $\Delta T = 660 K - 320 K$ is well described by $\eta_{\rm max}( Z_\gen^{(0)}, \tau^{(0)}, \beta^{(0)} | \gamma)$ over the whole $\gamma$ range for power generation; see \S10 and Figure S5 in SI.
Moreover, in high-performance materials, higher $Z_\gen^{(0)}$ usually implies higher thermoelectric efficiency even for segmented materials. Hence the top-ranked devices in order of their efficiency can be screened out quickly by using $Z_\gen^{(0)}$ only.
As an example, the efficiencies of 5-stage segmented devices comprised of 18 candidate materials are computed. Of nearly two million combinations of the segmented legs ($18^5 = 1,889,568$), a top 1\% high-$Z_\gen^{(0)}$ device is also a top 1\% high-efficiency device with 82\% probability; see \S12 in SI.

\subsection*{Application}
Our finding gives new direction for evaluation of thermoelectric materials and the design of thermoelectric devices, beyond $zT$. When $Z_\gen$ is modest, higher $\tau$ and $\beta$ may increase the thermoelectric performance.
For example, while the single crystalline SnSe has the highest peak$zT$ of 2.6 \cite{zhao2014ultralow}, 
Bi$_2$Te$_3$-based material has higher maximum efficiency.
This is because Bi$_2$Te$_3$ has a much higher $\tau$ than SnSe, due to its temperature-dependent material properties; see \S13 in SI for details.

Modulation of the thermoelectric degrees of freedom in functionally graded or segmented devices results in enhanced designs.
For example, we can design high-efficiency graded devices using Bi$_2$Te$_3$ by finding the optimal carrier concentration of each segmented region; see \S14 in SI.
We can also explore the efficiency space of segmented devices to check the current status and the limit of thermoelectric conversion.
We achieve a theoretical maximum efficiency $\eta_{\rm max}$ of 22.4\% at $\Delta T = 600 K$ using leg segmentation; see Figure \ref{fig-Ztaubeta-manage}(c).
Using the segmented legs, the efficiency is highly enhanced, up to 176\%, when $\Delta T > 600 K$.

\subsection*{Conclusion}
Three degrees of freedom in thermoelectrics, $Z_\gen$, $\tau$, and $\beta$, completely determine thermoelectric conversion efficiency, and they can be easily estimated from material properties. Each degree of freedom is a figure of merit, so improving one is a new way to increase efficiency. 
Modulating the thermoelectric degrees of freedom in segmented materials and legs, the efficiency can be enhanced up to 176\% compared to single-material devices.

\section*{Acknowledgement}
This work was supproted by Korea Electrotechnology Research Institute (KERI) Primary research program through the National Research Council of Science and Technology (NST) funded by the Ministry of Science and	 ICT (MSIT) of the Republic of Korea (No. 18-12-N0101-34) and partially supported by the Korea Institute of Energy Technology Evaluation and Planning (KETEP) and the Ministry of Trade, Industry and Energy (MOTIE) of the Republic of Korea (No. 20162000000910 and No. 20172010000830).

\section*{Contributions}
$\dagger$ B.R. and J.C. contributed equally to this work.
B.R. and S.D.P. designed the project. S.D.P. helped supervise the the project.
B.R. and J.C. developed the theoretical formalism, designed the computational framework, performed the calculations, analysed the results.
All authors, B.R., J.C. and S.D.P. discussed the results. B.R. and J.C. wrote the paper.

\section*{Competing interests}
The authors declare no competing financial interests.

\section*{Corresponding Author}
Correspondence to Byungki Ryu (byungkiryu@keri.re.kr) or Jaywan Chung. (jchung@keri.re.kr).

\section*{Methods}

\subsection*{Thermoelectric properties data}
Three thermoelectric properties (TEPs), Seebeck coefficient, electrical resistivity, and thermal conductivity, of 276 materials are gathered and digitised from 264 literatures.
In each reference, one sample, having the highest $zT$ value, is chosen for each type of material. 
If there are both $p$- and $n$-type materials, we sampled two TEP datasets.
The constructed thermoelectric property database consists of various material groups based on 
%\textcolor{red}{Bi$_2$Te$_3$, PbTe, SnSe, GeTe, AgSbTe$_2$, Mg$_2$Si, Si$_{1-x}$Ge$_{x}$, half-Heusler, skutterudite (SKD), and so on.}
Bi$_2$Te$_3$, PbTe, SnSe, GeTe, AgSbTe$_2$, Mg$_2$Si, Si$_{1-x}$Ge$_{x}$, half-Heusler, skutterudite (SKD), and so on.
The available temperature range of materials is determined from the available data range for TEPs, and linear interpolation is used to evaluate TEPs at arbitrary temperatures not given in the literatures.

\subsection*{Thermoelectric efficiency}
Using the temperature-dependent material properties, the thermoelectric efficiencies are calculated.
To describe the ideal maximum efficiency of a thermoelectric material, we assume a one-dimensional uni-leg, ignoring heat transfer by convection and radiation, as well as the non-ideal electrical and thermal losses occurred outside the material.
The temperature distribution $T(x)$ is determined by solving the differential equation modelling the thermoelectric effect with Dirichlet boundary conditions where, at the end points, the temperature is fixed by the available temperature range.
The thermoelectric performance is calculated as a function of electric current density $J$, given as
$\eta (J) = \frac{ J \left(   \int_{T_c}^{T_h} \alpha \,dT -  J \int_{T_c}^{T_h} \rho \,dx \right) }{ - (\kappa \frac{dT}{dx})_h + J \alpha_h T_h } $.
The maximum numerical efficiency $\eta_{\rm max}$ is searched so that the relation $\eta (J)  \leq \eta_{\rm max}$ holds.

%\begin{enumerate}
%\item Methods의 내용은 printed issue에는 나오지 않음.
%\item 간결하게 쓰되, 해석 및 결과물 재생성에 필요한 모든 elements를 쓸 것. (3천 단어 이하)
%\item 이미 출판된 방법론은 인용으로 충분.
%\item ``Data Availability'' statement 쓸 것.
%\item 그림이나 표는 방법론에 넣지 말 것.
%\item 항목마다 subsection을 이용해 설명할 것.
%\end{enumerate}

%\begin{thebibliography}{1}
%\bibitem{dummy} Articles are restricted to 50 references, Letters to 30.
%\bibitem{dummyb} No compound references -- only one source per reference.
%\end{thebibliography}

%\bibliographystyle{plain}
%\bibstyle{}

%\section*{References}
\bibliographystyle{naturemag}
\bibliography{ZT}

\end{document}

%\setpagewiselinenumbers
%\linenumbers

%\onecolumn
%\twocolumn

\maketitle

%\begin{affiliations}
% \item Energy Conversion Research Center, Korea Electrotechnology Research Institute (KERI), \newline 12, Bulmosanro 10beon-gil, Seongsan-gu, Changwon, 51543, Republic of Korea
%\end{affiliations}

%\author[Jaywan Chung]{Jaywan Chung}
%\author[Su-Dong Park]{Su-Dong Pak}
%\author[Byungki Ryu]{Byungki Ryu$^*$}

%\address[Byungki Ryu]{Thermoelectric Conversion Research Center (TCRC), Korea Electrotechnology Research Institute (KERI), \newline 12, Bulmosanro 10beon-gil, Seongsan-gu, Changwon, 51543, Republic of Korea}
%\email{byungkiryu@keri.re.kr}

%\date{\today}

%\keywords{thermoelectric, efficiency, figures of merit, temperature, heat}

\section{Thermoelectric Property Data used in the Manuscript}
In this work, we constructed a dataset of TEPs of 276 materials gathered from 264 literatures \cite{biswas_strained_2011, biswas_high-performance_2012, fu_realizing_2015, gelbstein_controlling_2013, he_ultrahigh_2015, heremans_enhancement_2008, hsu_cubic_2004, hu_shifting_2014, hu_power_2016, kim_dense_2015, lin_tellurium_2016, liu_thermoelectric_2011, liu_convergence_2012, pan_thermoelectric_2016, pei_convergence_2011, pei_high_2011-1, poudel_high-thermoelectric_2008, rhyee_peierls_2009, wang_right_2014, zhao_raising_2012, zhao_thermoelectrics_2012, zhao_ultralow_2014, zhao_ultrahigh_2015, cui_thermoelectric_2007, cui_crystal_2008, eum_transport_2015, fan_p-type_2010, han_alternative_2013, hsu_enhancing_2014, zheng_mechanically_2014, hu_tuning_2015, hwang_enhancing_2013, ko_nanograined_2013, zhang_improved_2015, zhao_bismuth_2005, lee_control_2010, lee_enhancement_2013, lee_crystal_2013, yan_experimental_2010, lee_preparation_2013, lee_preparation_2014, lee_preparation_2014-1, lee_preparation_2014-2, lee_thermoelectric_2014, sumithra_enhancement_2011, lukas_transport_2012, min_surfactant-free_2013, mun_fe-doping_2015, ovsyannikov_enhanced_2015, puneet_preferential_2013, shin_twin-driven_2014, son_n-type_2012, son_effect_2013, soni_enhanced_2012, xiao_enhanced_2014, tang_preparation_2007, wang_metal_2013, wu_thermoelectric_2013, yelgel_thermoelectric_2012, zhang_rational_2012, wei_minimum_2016, lan_high_2012, yu_preparation_2013, kosuga_enhanced_2014, scheele_thermoelectric_2011, ahn_improvement_2009, ahn_exploring_2010, ahn_enhanced_2013,androulakis_thermoelectric_2010, androulakis_high-temperature_2011, bali_thermoelectric_2013, bali_thermoelectric_2014, wu_superior_2015, dong_transport_2009, dow_effect_2010, falkenbach_thermoelectric_2013, fan_enhanced_2015, fang_synthesis_2013, jaworski_valence-band_2013, jian_significant_2015, keiber_complex_2013, kim_spinodally_2016, lee_improvement_2012, lee_contrasting_2014, li_enhanced_2013, li_pbte-based_2014, liu_effect_2013, lo_phonon_2012, lu_enhancement_2013, pei_combination_2011, pei_high_2011, pei_self-tuning_2011, pei_stabilizing_2011, pei_low_2012, pei_thermopower_2012, pei_optimum_2014, poudeu_high_2006, rawat_thermoelectric_2013, wang_large_2013, wang_tuning_2014, wu_broad_2014, wu_strong_2014, yamini_heterogeneous_2015, yang_enhanced_2015, zebarjadi_power_2011, zhang_enhancement_2012, zhang_heavy_2012, zhang_effect_2013, zhang_enhancement_2015, al_rahal_al_orabi_band_2015, banik_mg_2015, banik_high_2016, banik_agi_nodate, chen_thermoelectric_2014, chen_understanding_2016, leng_thermoelectric_2016, pei_interstitial_2016, tan_high_2014, tan_codoping_2015, tan_valence_2015, tang_realizing_2016, wang_thermoelectric_2015, zhang_high_2013, zhou_optimization_2014, guan_thermoelectric_2015, suzuki_supercell_2015, fahrnbauer_high_2015, gelbstein_-doped_2007, gelbstein_powder_2007, gelbstein_thermoelectric_2010, hazan_effective_2015, kusz_structure_2016, lee_influence_2014, schroder_nanostructures_2014, schroder_tags-related_2014, williams_enhanced_2015, wu_origin_2014, aikebaier_effect_2010, chen_thermoelectric_2012, dow_thermoelectric_2009, drymiotis_enhanced_2013, du_effect_2014, guin_sb_2015, han_lead-free_2012, he_synthesis_2012, hong_anomalous_2014, liu_enhanced_2016, mohanraman_influence_2014, pei_alloying_2011, wang_synthesis_2008, wu_state_2015, zhang_improved_2010, aizawa_solid_2006, akasaka_composition_2007, akasaka_non-wetting_2007, cheng_mg2si-based_2016, duan_effects_2016, isoda_effect_2007, kajikawa_thermoelectric_1998, liu_n-type_2015, luo_fabrication_2009, mars_thermoelectric_2009, noda_temperature_1992, tani_thermoelectric_2005, tani_thermoelectric_2007-1, tani_thermoelectric_2007, yang_preparation_2009, yin_optimization_2016, zhang_high_2008, zhang_situ_2008, zhang_suppressing_2015, zhao_synthesis_2009, joshi_enhanced_2008, tang_holey_2010, wang_enhanced_2008, ahn_improvement_2012, bhatt_thermoelectric_2014, fu_band_2015, kraemer_high_2015, krez_long-term_2015, liu_thermoelectric_2007, mudryk_thermoelectricity_2002, shi_low_2008, bai_enhanced_2009, bao_effect_2009, chitroub_thermoelectric_2009, dong_hpht_2009, duan_synthesis_2012, dyck_thermoelectric_2002, he_thermoelectric_2007, he_great_2008, laufek_synthesis_2009, li_thermoelectric_2005, liang_ultra-fast_2014, liu_enhanced_2007, mallik_transport_2008, mallik_thermoelectric_2013, mi_thermoelectric_2008, pei_thermoelectric_2008, qiu_high-temperature_2011, rogl_thermoelectric_2010, rogl_new_2011, rogl_n-type_2014, rogl_new_2015, sales_filled_1996, shi_multiple-filled_2011, stiewe_nanostructured_2005, su_structure_2011, tang_synthesis_2001, xu_thermoelectric_2014, yang_synthesis_2009, zhang_situ_2008-1, zhang_high-pressure_2012, zhao_enhanced_2009, zhou_thermoelectric_2013, bali_thermoelectric_2016, ding_high_2016, jo_simultaneous_2016, joo_thermoelectric_2016, li_enhanced_2016, li_inhibition_2016, liu_enhanced_2012, zhou_strategy_2017, zhou_scalable_2017, zhang_discovery_2017, xu_nanocomposites_2017, xie_stabilization_2016, wang_high_2016, seo_effect_2017, pei_multiple_2016, park_extraordinary_2016, moon_tunable_2016, zhu_nanostructuring_2007, choi_thermoelectric_1997, yamanaka_thermoelectric_2003, yang_nanostructures_2008, yang_natural_2010, zhang_effects_2009, zhou_nanostructured_2008, sharp_properties_2003, salvador_transport_2009, levin_analysis_2011, yamanaka_thermoelectric_2003-1, zhao_thermoelectric_2008, zhao_synthesis_2006, yu_high-performance_2009, xiong_high_2010, toberer_traversing_2008, chung_csbi4te6:_2000, tang_high_2008, mi_improved_2007, liu_improvement_2008, li_preparation_2008, chen_high_2006, zhong_high_2014, yu_thermoelectric_2012, liu_ultrahigh_2013, liu_copper_2012, he_high_2015, gahtori_giant_2015, day_high-temperature_2014, ballikaya_thermoelectric_2013, bailey_enhanced_2016, li_promoting_2017} to test our method.
The TEPs were digitized using the Plot Digitizer \cite{PlotDigitizer}.
The dataset consists of Seebeck coefficient $\alpha$, electrical resistivity $\rho$, and thermal conductivity $\kappa$ at measured temperature $T$.  For the numerical computation of efficiency, we use the available temperature ranges of the given material: the $T_c$ is defined as the maximum of the lowest measured temperautre and $T_h$ is defined as the minimum of the highest measured temperature for given materials.

As shown in Table \ref{tep-dataset}, 
the 276 materials in our dataset have various base-material groups: 59 $\rm Bi_2Te_3$-related materials, 55 $\rm PbTe$-related materials , 40 skutterudite (SKD), 23 $\rm Mg_2Si$-based materials, 18 $\rm GeTe$ materials, 14 $\rm M_2Q$ antifluorite-type chalcogenide materials (where M = Cu, Ag, Au and Q = Te, Se), 12 $\rm SnTe$-related materials, 11 $\rm ABQ_2$-type materials (where A=Group I, B=Bi, Sb, Q=Te, Se), 8 $\rm SnSe$-related materials, 7 $\rm PbSe$-related materials, 7 half-Heusler (HH) materials, 6 $\rm SiGe$-related materials, 3 $\rm In_4 Se_3$-related materials, 3 $\rm PbS$-related materials, 2 oxide materials, 2 clathrate materials, and 6 others. Here the base-material denotes the representative material, not the exact composition. Also note that for the categorizatoin of base materials, the doping element is ignored. For examples, $\rm Bi_2Te_3$, $\rm Sb_2Te_3$, $\rm Bi_2Se_3$ binary and their ternary alloys are categorized as $\rm Bi_2 Te_3$-related materials. The material doping composition is not denoted in the composition of the base material.

\begin{table}[h]
\caption{TEP Dataset of 276 materials with various material groups. `Group' and `\#mat.' coloumns represent the group of base material and the number of materials inside the Group.}
\label{tep-dataset}
\begin{tabular}{|c|c|c|c|}
\hline
\textbf{Group} & \textbf{\#mat.} & \textbf{Group} & \textbf{\#mat.} \\ \hline
$\rm Bi_2Te_3$ & 59 & SnSe & 8 \\ \hline
PbTe & 55 & PbSe & 7 \\ \hline
SKD & 40 & HH & 7 \\ \hline
$\rm Mg_2Si$ & 23 & SiGe & 6 \\ \hline
GeTe & 18 & $\rm In_4Se_3$ & 3 \\ \hline
$\rm M_2Q$ & 14 & PbS & 3 \\ \hline
SnTe & 12 & Oxide & 2 \\ \hline
$\rm ABQ_2$ & 11 & clathrate & 2 \\ \hline
etc. & 6 & \textbf{Total} & \textbf{276} \\ \hline
\end{tabular}
\end{table}

\begin{figure}
\centering \includegraphics[width=\textwidth]{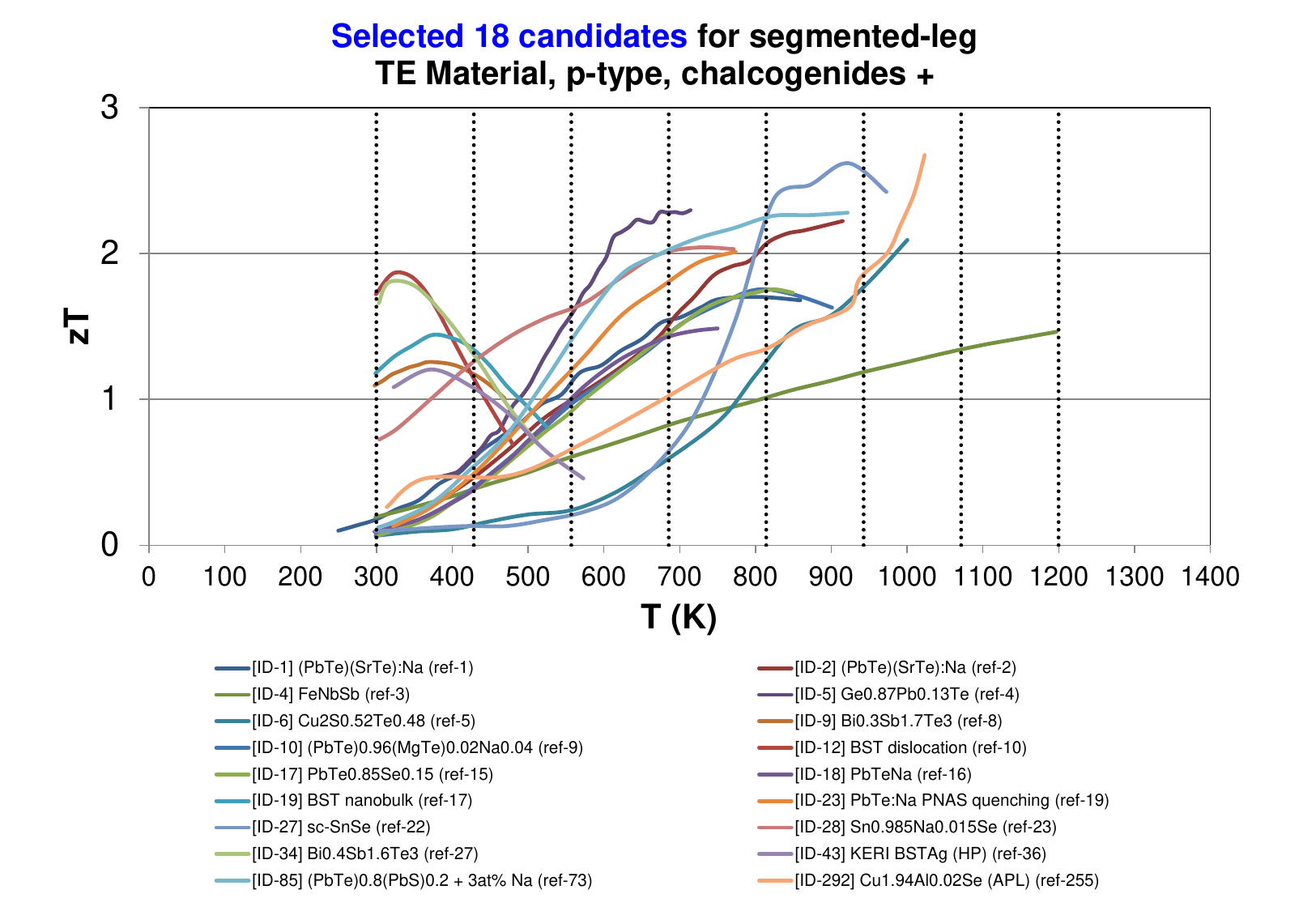}
\caption{The $zT$ curves for 18 selected materials. The `ref-\#' is the reference number.}\label{zT-for-18-mats}
\end{figure}

For the segmented-leg devices, we consider 18 candidates showing high peak $zT$ values exceeding 1. The full $zT$ curves of them are shown in Figure \ref{zT-for-18-mats}.
Table \ref{table-18-mats-temp-range}, \ref{table-18-mats-max-eff} and \ref{table-18-mats-te-dof} contain more information of the materials, including available temperature range, peak $zT$, numerical efficiency, formula efficiency, and the thermoelectric degrees of freedom.

\begin{table}[h]
\caption{Information of 18 selected materials: available temperature range $T_c$ and $T_h$, $\Delta T = T_h - T_c$, peak $zT$, temperature of the peak $zT$.}\label{table-18-mats-temp-range}
\begin{tabular}{|l|l|l|l|l|}
\hline
ID-\# & Material or Process [Reference] & $T_c$ (K) & $T_h$ (K) & peat-$zT$ @$T$ \\ \hline
ID-1 & (PbTe)(SrTe):Na  \cite{biswas_strained_2011} & 251 & 818 & 1.7  @800K \\ \hline
ID-2 & (PbTe)(SrTe):Na  \cite{biswas_high-performance_2012} & 302 & 915 & 2.2  @915K \\ \hline
ID-4 & FeNbSb  \cite{fu_realizing_2015} & 301 & 1200 & 1.5  @1200K \\ \hline
ID-5 & $\rm Ge_{0.87}Pb_{0.13}Te$  \cite{gelbstein_controlling_2013} & 329 & 713 & 2  @673K \\ \hline
ID-6 & $\rm Cu_2S_{0.52}Te_{0.48}$  \cite{he_ultrahigh_2015} & 299 & 997 & 2.1  @1000K \\ \hline
ID-9 & $\rm Bi_{0.3}Sb_{1.7}Te_3$  \cite{hu_shifting_2014} & 298 & 479 & 1.3  @380K \\ \hline
ID-10 & $\rm (PbTe)_{0.96}(MgTe)_{0.02}Na_{0.04}$  \cite{hu_power_2016} & 307 & 900 & 1.8  @810K \\ \hline
ID-12 & BST  dislocation \cite{kim_dense_2015} & 300 & 480 & 1.86  @320K \\ \hline
ID-17 & $\rm PbTe_{0.85}Se_{0.15}$  \cite{pei_convergence_2011} & 300 & 847 & 1.8  @850K \\ \hline
ID-18 & PbTeNa  \cite{pei_high_2011-1} & 300 & 750 & 1.4  @750K \\ \hline
ID-19 & BST nanobulk  \cite{poudel_high-thermoelectric_2008} & 300 & 525 & 1.4  @373K \\ \hline
ID-23 & PbTe:Na, quenching (PNAS) \cite{wang_right_2014} & 321 & 759 & 2  @773K \\ \hline
ID-27 & sc-SnSe, $b$-axis  \cite{zhao_ultralow_2014} & 303 & 970 & 2.6  @923K \\ \hline
ID-28 & $\rm Sn_{0.985}Na_{0.015}Se$  \cite{zhao_ultrahigh_2015} & 304 & 773 & 2  @773K \\ \hline
ID-34 & $\rm Bi_{0.4}Sb_{1.6}Te_3$  \cite{fan_p-type_2010} & 303 & 513 & 1.8  @316K \\ \hline
ID-43 & KERI BSTAg, HP  \cite{lee_control_2010} & 323 & 573 & 1.2  @373K \\ \hline
ID-85 & $\rm (PbTe)_{0.8}(PbS)_{0.2}$ + 3at\% Na \cite{wu_superior_2015} & 302 & 922 & 2.3  @923K \\ \hline
ID-292 & $\rm Cu_{1.94}Al_{0.02}Se$ (APL)  \cite{zhong_high_2014} & 327 & 1019 & 2.62  @1029K \\ \hline
\end{tabular}
\end{table}

\begin{table}[h]
\caption{Information of 18 selected materials: (a) maximum efficiencies computed using exact numerical method ($T$ is computed by fixed-point interation, then power, heat and efficiency are computed), 
maximum efficiencies computed from general maximum efficiency formula $\eta_{\rm max}^{\rm gen}$ (see equation \eqref{SI-max-efficiency}) 
(b) using \emph{exact} thermoeletric degrees of freedom (DOFs) with exact $T$ ($Z_\gen,\tau,\beta$), 
(c) using DOFs with $T^{(0)}$ ($Z_\gen^{(0)},\tau^{(0)},\beta^{(0)}$), 
(d) using DOFs with \emph{one-shot} approximation ($Z_\gen^{(0)},\tau_{\rm lin}^{(0)},\beta_{\rm lin}^{(0)}$),
(e) using DOFs with only $Z_\gen$ while $\tau=\beta=0$,
(f) using DOFs with only $Z_\gen^{(0)}$ while $\tau=\beta=0$,
and (g) using the classical efficiency formula for constant TEP using peak $zT$.
Note that when we compute the numerical maximum efficiency we calculate the $T$ using the fixed-point iteration with integral equation of $T$ for given $J$. Then $J$ is optimized to maximize the efficiency.
Note that when we use the general maximum efficiency formula, the $T$ and $J$ are simultaneously computed. For $T$, the fixed-point iteration is used. For $J$, we use the optimal $\gamma$ formula $\gamma_{\rm max}^{\rm gen}$.
}
\label{table-18-mats-max-eff}
\begin{tabular}{|l|c|c|c|c|c|c|c|}
\hline
\multirow{3}{*}{ID-\#} & \multicolumn{7}{c|}{$\eta_{\rm max}$} \\ \cline{2-8} 
 & \multirow{2}{*}{\makecell{(a)\\ exact}} & \multicolumn{5}{c|}{$\eta_{\rm max}^{\rm gen}$} 
 &  $\eta_{\rm max}^{\rm const}$ \\ \cline{3-8} 
 &  
 & \makecell{(b)\\ $Z_\gen,\tau,\beta$}
 & \makecell{(c)\\ $Z_\gen^{(0)},\tau^{(0)},\beta^{(0)}$} 
 & \makecell{(d)\\ $Z_\gen^{(0)},\tau_{\rm lin}^{(0)},\beta_{\rm lin}^{(0)}$}
 & \makecell{(e)\\ $Z_\gen$}
 & \makecell{(f)\\ $Z_\gen^{(0)}$}
 & \makecell{(g)\\ peak $zT$} \\ \hline
ID-1 & 13.7\% & 13.7\% & 14.4\% & 14.3\% & 14.5\% & 15\% & 22.9\% \\ \hline
ID-2 & 15.9\% & 15.9\% & 16.2\% & 16.1\% & 16.6\% & 16.8\% & 24.9\% \\ \hline
ID-4 & 15.3\% & 15.3\% & 15.8\% & 15.8\% & 15.8\% & 16.3\% & 23.8\% \\ \hline
ID-5 & 12.5\% & 12.6\% & 12.9\% & 13\% & 13.1\% & 13.4\% & 18\% \\ \hline
ID-6 & 10.5\% & 10.5\% & 10.7\% & 10.7\% & 11.1\% & 11.1\% & 25.9\% \\ \hline
ID-9 & 8.4\% & 8.4\% & 8.4\% & 8.4\% & 8.4\% & 8.4\% & 9.2\% \\ \hline
ID-10 & 13.8\% & 13.8\% & 14.2\% & 14.1\% & 14.4\% & 14.7\% & 22\% \\ \hline
ID-12 & 9.1\% & 9.1\% & 9.1\% & 9.1\% & 9\% & 9\% & 11.2\% \\ \hline
ID-17 & 12.6\% & 12.7\% & 13\% & 12.9\% & 13.3\% & 13.5\% & 21.5\% \\ \hline
ID-18 & 10.4\% & 10.4\% & 10.8\% & 10.8\% & 10.9\% & 11.2\% & 16.9\% \\ \hline
ID-19 & 9.9\% & 9.9\% & 10\% & 10\% & 9.9\% & 9.9\% & 11.1\% \\ \hline
ID-23 & 11.6\% & 11.6\% & 12.1\% & 12.1\% & 12.2\% & 12.5\% & 19.6\% \\ \hline
ID-27 & 7.1\% & 7.1\% & 7.1\% & 7.1\% & 7.1\% & 7.1\% & 27.9\% \\ \hline
ID-28 & 16.2\% & 16.2\% & 16.9\% & 16.9\% & 16.7\% & 17.3\% & 20.9\% \\ \hline
ID-34 & 10.1\% & 10.1\% & 10.1\% & 10.1\% & 10\% & 10\% & 12.2\% \\ \hline
ID-43 & 8.2\% & 8.2\% & 8.2\% & 8.2\% & 8.1\% & 8.1\% & 10.3\% \\ \hline
ID-85 & 17.6\% & 17.6\% & 18.1\% & 17.8\% & 18.5\% & 18.8\% & 25.6\% \\ \hline
ID-292 & 14.3\% & 14.3\% & 14.9\% & 14.9\% & 14.9\% & 15.4\% & 27.5\% \\ \hline
\end{tabular}
\end{table}

\begin{table}[h]
\caption{Information of 18 selected materials: \emph{exact} value and \emph{one-shot} approximation of thermoeletric degrees of freedom.}
\label{table-18-mats-te-dof}
\begin{tabular}{|l|r|r|r|r|r|r|}
\hline
\multicolumn{1}{|c|}{ID-\#} & \multicolumn{1}{c|}{$Z_\gen$} & \multicolumn{1}{c|}{$\tau$} & \multicolumn{1}{c|}{$\beta$} & \multicolumn{1}{c|}{$Z_\gen^{(0)}$} & \multicolumn{1}{c|}{$\tau_{\rm lin}^{(0)}$} & \multicolumn{1}{c|}{$\beta_{\rm lin}^{(0)}$} \\ \hline
ID-1 & 0.0015 & -0.253 & 0.192 & 0.0016 & -0.207 & 0.199 \\ \hline
ID-2 & 0.0018 & -0.186 & 0.068 & 0.0018 & -0.152 & 0.074 \\ \hline
ID-4 & 0.0010 & -0.164 & 0.197 & 0.0011 & -0.141 & 0.203 \\ \hline
ID-5 & 0.0022 & -0.227 & 0.094 & 0.0023 & -0.168 & 0.105 \\ \hline
ID-6 & 0.0008 & -0.253 & 0.027 & 0.0008 & -0.208 & 0.028 \\ \hline
ID-9 & 0.0029 & -0.019 & 0.135 & 0.0029 & -0.017 & 0.136 \\ \hline
ID-10 & 0.0015 & -0.192 & 0.102 & 0.0015 & -0.161 & 0.107 \\ \hline
ID-12 & 0.0033 & 0.030 & 0.177 & 0.0033 & 0.032 & 0.178 \\ \hline
ID-17 & 0.0014 & -0.231 & 0.109 & 0.0015 & -0.189 & 0.112 \\ \hline
ID-18 & 0.0014 & -0.271 & 0.167 & 0.0014 & -0.214 & 0.172 \\ \hline
ID-19 & 0.0028 & -0.015 & 0.189 & 0.0028 & -0.013 & 0.190 \\ \hline
ID-23 & 0.0017 & -0.254 & 0.138 & 0.0017 & -0.194 & 0.142 \\ \hline
ID-27 & 0.0005 & 0.082 & -0.379 & 0.0005 & 0.086 & -0.382 \\ \hline
ID-28 & 0.0025 & -0.154 & 0.217 & 0.0026 & -0.118 & 0.225 \\ \hline
ID-34 & 0.0032 & 0.033 & 0.164 & 0.0032 & 0.036 & 0.166 \\ \hline
ID-43 & 0.0019 & 0.028 & 0.186 & 0.0019 & 0.029 & 0.187 \\ \hline
ID-85 & 0.0021 & -0.179 & 0.079 & 0.0021 & -0.146 & 0.095 \\ \hline
ID-292 & 0.0013 & -0.211 & 0.178 & 0.0014 & -0.166 & 0.187 \\ \hline
\end{tabular}
\end{table}

\clearpage
\section{Numerical Efficiency Calculation in Figure 1}
Numerical maximum efficiencies of ideal thermoelectric devices without thermal loss by radiation or air convection are computed for 276 materials and compared with the peak $zT$ values.
The thermoelectric properties are \emph{linearly interpolated} at intermediate temperatures.
The exact temperature distribution $T(x)$ of steady state is determined by solving the differential equations of thermoelectricity with Dirichlet boundary conditions; the end point temperature is determined from the available temperature range.
Then the thermoelectric performances of a thermoelectric leg with length $L$ and cross sectional area $A$ are calculated as a function of current density $J$ given as
$\eta (J) = \frac{P/A}{Q_h/A} = \frac{ J (   \int_c^h \alpha dT -  J \int_0^L \rho dx ) }{ -   \kappa_h \nabla T_h + J \alpha_h T_h } $, where the $P$ and $Q_h$ are the power delivered outside and the hot-side heat current respectively. Then, the maximum of numerical efficiency ($\eta_{\rm max}$) is calculated, which satisfies the relation $\eta (J)  \leq \eta_{\rm max}$.
The reduced efficiency $\eta_{\rm red}$ is obtained as $\eta_{\rm red} = \frac{\eta_{\rm max}}{\eta_{\rm Carnot}}$, where $\eta_{\rm Carnot} = \frac{T_h - T_c}{T_h}$.

\section{Device Parameters and Operating Conditions}

The thermoelectric (TE) power device mentioned in this paper is a uni-leg device composed of a single leg or a segmented leg sandwiched by heat source ($T_h$) and heat sink ($T_c$) at both sides. 
In such a device, electric current and heat current flow simultaneously across the leg. For the simplicity, we assume the steady-state condition.
For $p$-type material ($\alpha > 0$), the electric current and the heat current flow in the same direction from hot to cold side, 
while the direction of the electric current is reversed in $n$-type material ($\alpha < 0$).

\begin{figure}
\centering \includegraphics[width=0.8\textwidth]{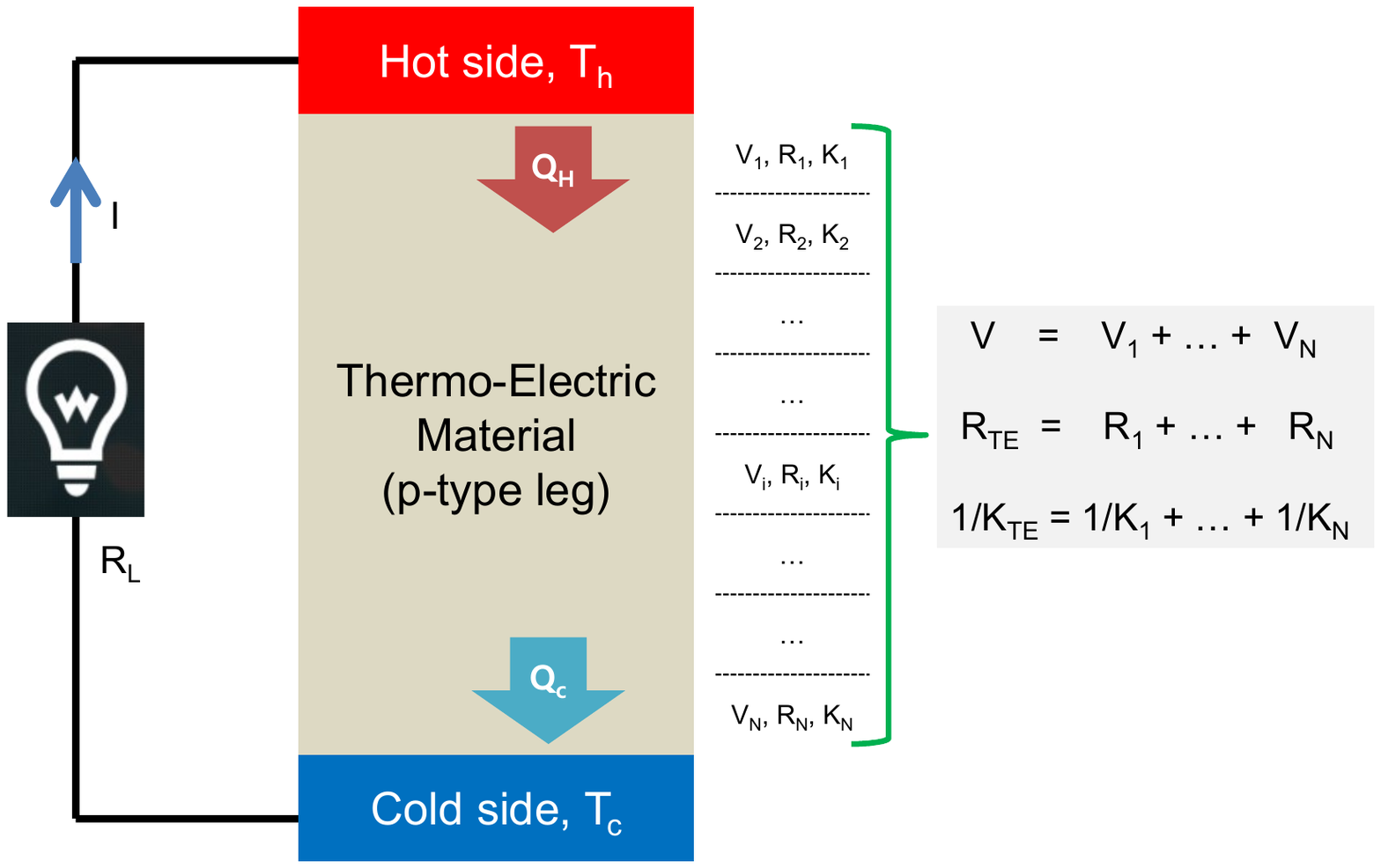}
\caption{Structure of conventional thermoelectric power devices. For simplicity, we draw only an uni-leg with $p$-type materials where electric current flows from hot to cold side. Since the electric current and heat current flow though the leg, the electrical and thermal resistance of the leg should be considered as the sum of an infinitesimal serial circuit. Thus, the voltage $V$ and the resistance $R$ should be the sum of component voltages and resistances respectively. In the case of thermal conduction, the inverse of thermal conductivity should be used for thermal circuit parameter.}
\label{fig-deviceParameter}
\end{figure}

The most important parameters in a TE device are voltage $V$, electrical resistance $R$, and thermal resistance $1/K$, which can describe the electrical and thermal circuits of the TE device.  
Once these three device parameters are known, we can roughly estimate the thermoelectric performance of the TE device. 
When there is load resistance $R_{\rm L}$, there will be electric current $I = \frac{V}{R+R_{\rm L}}$. 
When there is no electric current, there will be heat current $Q_h = - A \kappa  \nabla T = K \Delta T$. When there is non-zero electric current, there will be heat generation by Thomson and Joule heat and the hot side heat current will be approximately $Q_h \approx K \Delta T + I \frac{V}{\Delta T} T_h - \frac{1}{2} I^2 R$. The approximation becomes exact when there is no temperature dependency in thermoelectric properties (TEPs).
The three parameters $V,R,K$ are easily determined from the TE properties. Note that a leg of the device is equivalent to a series of infinitesimal parts $dx$, and it is trivial to write the induced open-circuit voltage ($V$) as an integration of $-\alpha \nabla T$ on $x$, and the resistance of the TE leg ($R_{TE}$ and $1/ K_{TE}$) as an integration of resistivity {$\rho$ and $1/ \kappa$} on $x$; see Figure \ref{fig-deviceParameter}. Also note that the electrical and thermal resistances should be calculated by integration of the corresponding resistivities on $x$, not on $T$.

When the material thermoelectric figure of merit $zT$ is small, the electric current density $J$ is so small that the $R$ and $K$ can be estimated by $R^{(0)}$ and $K^{(0)}$ which are the electrical resistance and thermal conductance for zero-current-density case ($J=0$).
Similarly, since the $J$ is small, the temperature can be estimated by the zero-current-density solution $T^{(0)} (x)$ which is the solution of the heat equation $\nabla \cdot (\kappa \nabla T) = 0$ without thermoelectric heat generation. Here the $\kappa$ is thermal conductivity.
The heat flows are nearly the same along the thermoelectric leg so the one-dimensional heat equation suggests $\kappa \frac{dT}{dx}$ is constant. Hence the average thermal conductivity $\kappaBar^{(0)}$ for $J=0$ satisfies $\kappaBar^{(0)} \frac{\Delta T}{L} = \kappa \frac{dT}{dx}$ so it can be evaluated by integration over $T$: $\kappaBar^{(0)} = \int \kappaBar^{(0)} \frac{1}{L} dx = \frac{1}{\Delta T} \int \kappa \frac{dT}{dx} dx  = \langle \kappa \rangle_T$ by the change of variable $dx= \frac{\kappa dT}{\kappaBar_0 \frac{\Delta T}{L}}$. Here the $\langle \kappa \rangle_T$ denotes the average of the thermal conductivity $\kappa(T)$ over $T$.
Meanwhile, the resitivity under the condition of $J=0$ is calculated as 
$\rhoBar^{(0)} = \frac{1}{L} \int \rho dx = \frac{1}{L} \int \rho \frac{\kappa dT}{\kappaBar^{(0)} \frac{\Delta T}{L}} = \frac{1}{\kappaBar^{(0)} \Delta T} \int \rho \kappa dT = \frac{\left< \rho \kappa \right>_T}{\left< \kappa \right>_T}$.
Finally we may rewrite $RK = \rhoBar \,\kappaBar \approx \rhoBar^{(0)} \kappaBar^{(0)} = \left< \rho \kappa \right>_T$ under small $zT$.

The above idea to use the device parameters for $J=0$ is the main idea of the one-shot approximation, of which argument is dealt thoroughly in $\S$\ref{sec-one-shot}.
Every thermoelectric material at the moment has the peak $zT$ smaller than 3, implying that the above idea gives a good approximation $Z_{\rm gen}^{(0)}$ for $Z_{\rm gen}$; see \eqref{one-shot-Zgen} for its definition.
However, under large $zT$ or non-zero $J$, the approximation $Z_{\rm gen}^{(0)}$ may have 1 to 10 percent error.

\section{Thermoelectric differential equation in one-dimension}
The thermoelectric effect is expressed in terms of electric current density $J$ and heat current density $J^Q$: $J = \sigma ( E - \alpha \nabla T)$ and $J^Q =  \alpha T J - \kappa \nabla T$ where $E$ is electric field. Applying the energy conservation law on $J$ and $J^Q$ and assuming \emph{one-dimensional} circuit case, we can obtain the thermoelectric differential equation\cite{chung2014nonlocal, goupil2015continuum} describing evolution of temperature distribution $T(x)$ inside an uni-leg thermoelectric device:
\begin{equation}\label{SI-TEQ-1D}
\frac{d}{dx} \left(\kappa \frac{dT}{dx} \right) + \rho J^2 -T \frac{d \alpha}{dT} \frac{dT}{dx} J = 0
\end{equation}
where $x$ is coordinate inside the one-dimensional thermoelectric leg. We have Dirichlet boundary condition since the temperatures at the end of the leg are fixed:
\begin{equation}\label{SI-TEQ-BC}
T(0) = T_h, \quad T(L) = T_c.
\end{equation}
In the one-dimensonal leg, where the cross sectional area $A$ is constant across the leg, the electric current is calculated as $I = J \times A $ 
and the heat current is calculated as $Q = J^Q \times A$

\section{Average Parameters and general figure of merit $Z_\gen$}
To analyze the thermoelectric equation \eqref{SI-TEQ-1D}, the following average material properties are helpful:
\begin{alignat*}{3}
\alphaBar &:=  \frac{1}{\Delta T} \int_{T_c}^{T_h}  \alpha \,dT &=& \frac{V}{\Delta T}, \\
\rhoBar &:= \frac{1}{L} \int_{0}^{L} \rho \,dx &=& \frac{A}{L} R, \\
\frac{1}{\kappaBar} &:= \frac{1}{L} \int_{0}^{L} \frac{1}{\kappa} \,dx &=& \frac{A}{L} \frac{1}{K}.
\end{alignat*}
Note the average parameters give the induced open-circuit voltage $V$, electrical resistance $R$ and thermal resistance $1/K$ of the leg.
Using these parameters we also define the \emph{general device figure of merit} $Z_{\gen}$ for temperature dependent material properties:
\begin{equation} \label{def-Zgen}
Z_{\gen} := \frac{\alphaBar^2}{R K} =  \frac{\alphaBar^2}{\rhoBar \,\kappaBar},
\end{equation}
which generalize the classical device figure of merit. If the material properties are temperature independent, the $Z_{\gen}$ is reduced to the conventional material parameter $z$.

\section{Electric Current Equation}
With given load resistance $R_{\rm L}$, an equation for the electric current density $J = \sigma \left(E - \alpha \frac{dT}{dx}\right)$ can be found by integrating $\rho J$ along the closed circuit: $\oint \rho J \,dx = \oint E \,dx - \oint \alpha \frac{dT}{dx}\,dx = V$. Hence the electric current $I$ satisfies $(R + R_{\rm L}) I = V$ and we have
\begin{equation}\label{SI-TEQ-J}
J = \frac{1}{A} \frac{V}{R+R_{\rm L}}.
\end{equation}
Note that the $R = \frac{1}{A} \int_0^L \rho \left( T(x) \right) \,dx$ depends on $T$ so does the $J$.

\section{Integral Equations of $T(x)$ and $\nabla T(x)$}
Due to the nonlinearity ($\kappa$, $\alpha$, $\rho$ depend on $T$) and nonlocality ($J$ depends on an integral of $T$) \cite{chung2014nonlocal}, the equation \eqref{SI-TEQ-1D} does not have an analytic solution. Instead, we rewrite the equation as an integral form where fixed-point iteration is possible. The integral equation will give us physical insights to derive the remaining degrees of freedom $\tau$ and $\beta$.

\begin{figure}
\centering\includegraphics[width=0.8\textwidth]{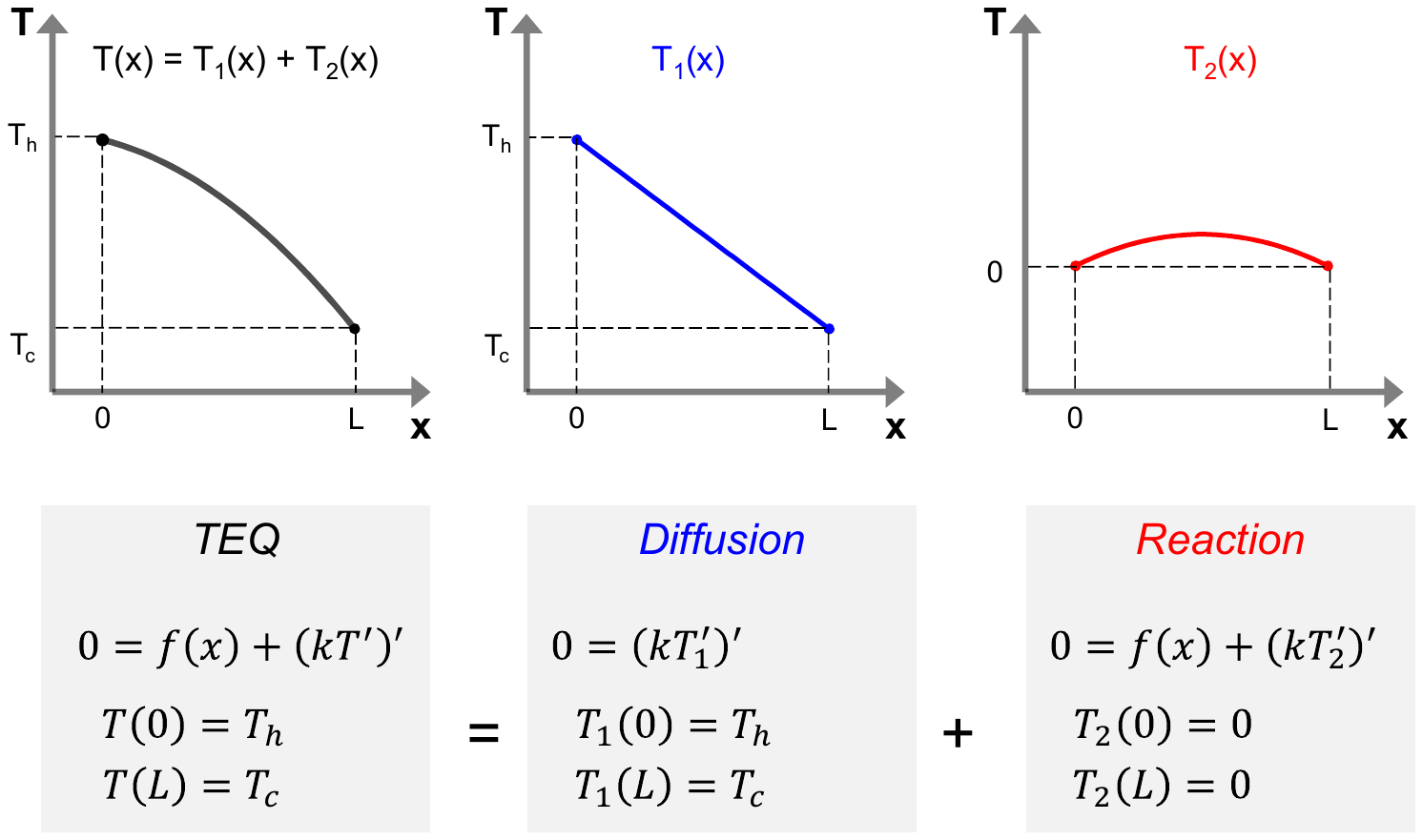}
\caption{Crucial idea to solve the temperature equation. The solution of the PDE \eqref{SI-TEQ-1D} can be decomposed to $T_1(x)$ and $T_2(x)$ with proper boundary conditions. Without the reaction term, the solution becomes simple, while still having physical meaning due to the relatively small contribution of $T_2(x)$ in thermoelectric legs.}\label{fig-tDecomposition}
\end{figure}

For simplicity, we denote the term with Joule heat and Thomson heat by $f_T(x)$:
\begin{equation} \label{SI-fT}
f_T(x) := \rho J^2 - T \frac{d \alpha}{dT} \frac{dT}{dx} J.
\end{equation}
Then the equation \eqref{SI-TEQ-1D} is $\frac{d}{dx} \left( \kappa \frac{dT}{dx} \right) + f_T = 0$. If the solution $T_{\rm sol}$ of \eqref{SI-TEQ-1D}, \eqref{SI-TEQ-BC}, \eqref{SI-TEQ-J} is known, we may put $\kappa(x) := \kappa(T_{\rm sol}(x))$ and $f(x) := f_{T_{\rm sol}}(x)$ to find a \emph{linear} equation
\begin{equation} \label{SI-linear-eq}
\frac{d}{dx} \left( \kappa(x) \frac{dT}{dx} \right) + f = 0.
\end{equation}
Since this equation is linear, we can find a solution by decomposing it into a homogeneous solution $T_1$ and particular solution $T_2$: $T = T_1 + T_2$. The $T_1$ and $T_2$ are solutions of
\begin{alignat}{6}
\frac{d}{dx} \left( \kappa(x) \frac{dT_1}{dx} \right) &=& 0, \quad T_1(0) &=& T_h, \quad T_1(L) &=& T_c, \label{SI-T1-eq} \\
\frac{d}{dx} \left( \kappa(x) \frac{dT_2}{dx} \right) + f &=& 0, \quad T_2(0) &=& 0, \quad T_2(L) &=& 0. \label{SI-T2-eq}
\end{alignat}
This idea is summarized in Figure \ref{fig-tDecomposition}.
To solve the equation \eqref{SI-T1-eq}, we integrate it over $x$ to yield $\kappa(x) \frac{dT_1}{dx}(x) = C$ for some constant $C$. Dividing both sides by $\kappa$ and integrating from $0$ to $x$, we have $T_1(x) - T_1(0) = C \int_0^x \frac{1}{\kappa(x)}\,dx$. Imposing the boundary conditions yields $C  = -K \frac{T_h - T_c}{A}$ and
\[ T_1 (x) = T_h - \frac{K \Delta T}{A} \int_0^x \frac{1}{\kappa(x)}\,dx. \]
To solve the equation \eqref{SI-T2-eq}, we integrate it from $0$ to $x$ to yield $\kappa(x) \frac{dT_2}{dx}(x) -C = -\int_0^x f(s)\,ds =: -F(x)$ for some constant $C$. Dividing both sides by $\kappa$ and integrating from $0$ to $x$, we have $T_2 (x) - T_2 (0) =  - \int_0^x \frac{F(x)}{\kappa (x)} \,dx + C \int_0^x \frac{1}{\kappa(x)} \,dx$. Imposing the zero boundary conditions yields
\[ T_2 (x) = -\int_0^x \frac{ F(x)}{\kappa(x)} \,dx +\frac{ K \,\delta T}{A} \int_0^x \frac{1}{\kappa(x)} \,dx, \]
where $\delta T := \int_0^L \frac{F(x)}{\kappa (x)} \,dx$ is a scalar quantity.

Summing up, we have the solution $T=T_1 + T_2$ of \eqref{SI-linear-eq}, \eqref{SI-TEQ-BC} and its gradient:
\begin{align}
T(x) &=  \left( T_h - \frac{K \Delta T}{A} \int_0^x \frac{1}{\kappa} \,dx  \right)
+  \left( -\int_0^x \frac{F(x)}{\kappa(x)}\,dx +\frac{ K \,\delta T}{A} \int_0^x \frac{1}{\kappa} \,dx \right)  \label{SI-T-integral-form} \\
\frac{dT}{dx}(x) &=  \left(  -\frac{K \Delta T}{A}  \frac{1}{\kappa (x)} \right)
+  \left( \frac{F(x)}{\kappa(x)} +\frac{ K \,\delta T}{A} \frac{1}{\kappa (x)}  \right) \label{SI-gradT}
\end{align}
where $F(x) = \int_0^x f(s)\,ds$ and $\delta T =  \int_0^L \frac{F(x)}{\kappa (x)} \,dx$. Since $\kappa(x)=\kappa(T(x))$ and $f(x) = f_{T}(x)$, the equation \eqref{SI-T-integral-form} is an integral form $T = \varphi[T]$ where $\varphi$ is the integral operator in the right-hand side of \eqref{SI-T-integral-form}.

To find $T$, we apply fixed-point iteration \cite{burden2010numerical} to the relation $T = \varphi[T]$. Choosing an initial guess $T_0$ for $T$ (it can be a linear distribution satisfying Dirichlet condition or the temperature curve satisfying $J=0$), we iteratively compute a sequence of functions $T_{n+1} = \varphi[T_n]$ for $n\geq 0$. Then we expect $T_n$ converges to a function $T_\infty$ which is the solution we are looking for because it satisfies $T_\infty = \varphi[T_\infty]$. Computation reveals that with linear $T_0$, the $T_n$ converges enough within a few iterations (less than 10 iterations).

\section{Heat current and additional figure of merit $\tau$ and $\beta$}
Using the $\frac{dT}{dx}$ in \eqref{SI-gradT}, the hot-side heat current can be written as
\begin{equation} \label{SI-qHot}
Q_h = A J_h^Q = I \alpha_h T_h - A \kappa_h \Big(\frac{dT}{dx}\Big)_h = I \alpha_h T_h + K (\Delta T - \delta T).
\end{equation}
Now we decompose $\delta T$ into two terms having $I$ and $I^2$. From \eqref{SI-fT},
\[\begin{split}
F_T(x) = \int_0^x f_T(s)\,ds &= I^2 \int_0^x \frac{1}{A^2} \rho(s)\,ds - I \int_0^x \frac{1}{A} T(s) \frac{d \alpha}{dT}(T(s)) \frac{dT}{dx}(s) \,ds \\
&=: I^2 F_T^{(2)}(x) - I F_T^{(1)}(x).
\end{split}\]
Hence
\[\begin{split}
\delta T = \int_0^L \frac{F_T(x)}{\kappa(x)}\,dx &= I^2 \int_0^L \frac{F_T^{(2)}(x)}{\kappa(x)}\,dx - I \int_0^L \frac{F_T^{(1)}(x)}{\kappa(x)}\,dx \\
&=: I^2 \delta T^{(2)} - I \delta T^{(1)}.
\end{split}\]
For \emph{temperature-independent} material properties, we can easily check that $\delta T^{(2)} = \frac{1}{2}\frac{R}{K}$ and $\delta T^{(1)} \equiv 0$ so that the hot-side heat current is
\[\QhBar = K \Delta T + I \alphaBar T_h -\frac{1}{2} I^2 R.\]
Our strategy is to consider the $Q_h$ in \eqref{SI-qHot} as a perturbation of $\QhBar$ above. To do so, we replace $\alpha_h$ by $\alphaBar$ in \eqref{SI-qHot} and introduce dimensionless perturbation parameters $\tau$ and $\beta$ of which values become zero for temperature-independent material properties. Precisely we let
\begin{align}
\tau &:= \frac{1}{\alphaBar \Delta T} \left[ (\alphaBar - \alpha_h) T_h -K \,\delta T^{(1)} \right], \label{SI-tau}\\
\beta &:= \frac{2}{R} K \,\delta T^{(2)} - 1. \nonumber %\label{SI-beta}
\end{align}
Then we can rewrite the $Q_h$ in \eqref{SI-qHot} by
\begin{equation}\label{SI-Qh-tau-beta}
Q_h =  K \Delta T + I \alphaBar ( T_h - \tau \Delta T)  - \frac{1}{2} I^2 R (1+\beta).
\end{equation}
Observing the delivered power $P=I (V-IR)=I(\alphaBar \Delta T-IR)$ is equal to $Q_h-Q_c$, we have the cold-side heat current:
\begin{equation}\nonumber %\label{SI-Qc-tau-beta}
Q_c =  K \Delta T + I \alphaBar ( T_c - \tau \Delta T)  +\frac{1}{2} I^2 R (1-\beta).
\end{equation}
When the average device parameters are fixed, the $Q_h$ in \eqref{SI-Qh-tau-beta} decreases as $\tau$ or $\beta$ increases while the delivered power $P$ is fixed. Hence the efficiency $\eta = \frac{P}{Q_h}$ increases as $\tau$ or $\beta$ increases. This implies each of $\tau$ and $\beta$ is a figure of merit for efficiency, as well as $Z_\gen$ is.

\section{Efficiency prediction using thermoelectric degrees of freedom}
Here we derive an efficiency formula in terms of the thermoelectric degrees of freedom $Z_\gen$, $\tau$, $\beta$ and find the maximum efficiency.
Let $\gamma := \frac{R_{\rm L}}{R}$. Then the electric current is $I = \frac{\alphaBar \Delta T}{R (1+ \gamma)} $ and the delivered power is $P=I(\alphaBar \Delta T-IR) = \frac{(\alphaBar \Delta T)^2}{R} \frac{\gamma}{(1+ \gamma)^2}$. Using \eqref{SI-Qh-tau-beta}, the efficiency $\eta = \frac{P}{Q_h} = \frac{P/(K\Delta T)}{Q_h/(K\Delta T)}$ can be written as
\begin{equation} \nonumber % \label{SI-efficiency-tau-beta}
\eta (Z_\gen, \tau, \beta | T_h, T_c, \gamma) = \frac{  Z_\gen \Delta T  \frac{\gamma}{(1+ \gamma)^2 } }{
 1 + Z_\gen  \big( \frac{1}{1+ \gamma} \big) ( T_h - \tau \Delta T ) 
 -\frac{1}{2} Z_\gen \Delta T \big( \frac{1}{1+ \gamma} \big)^2 (1+\beta).
 } 
\end{equation}
We can easily check that the efficiency is monotonic on $Z_\gen$, $\tau$ and $\beta$ for fixed $T_h,T_c$ and $\gamma$. \emph{Assuming $Z_\gen$, $\tau$, $\beta$ changes little} near the $\gamma$ at the maximum efficiency, we solve $\frac{\partial \eta}{\partial \gamma} = 0$ to estimate the maximum efficiency. For simplicity, we let
\[ T_h' := T_h -\tau \Delta T, \quad T_c' := T_c - (\tau+\beta)\Delta T, \quad T_m' := \frac{1}{2}(T_h' +T_c').\]
Then the solution of $\frac{\partial \eta}{\partial \gamma} = 0$ is
\begin{equation}\label{gamma-gen-max}
\gamma_{\rm max}^{\rm gen} = \sqrt{ 1+Z_\gen T_m' }.
\end{equation}
Hence the maximum efficiency is approximated by
\begin{equation} \label{SI-max-efficiency}
\eta_{\rm max} \approx \eta_{\rm max}^{\rm gen}  := \frac{\Delta T}{T_h'} \frac{ \sqrt{1+Z_\gen T_m'}-1}{\sqrt{1+Z_\gen T_m'} +\frac{T_c'}{T_h'}}.
\end{equation}
This formula generalizes the classical maximum efficiency formula for temperature-independent material properties because it has the same form as the classical formula but predicts the exact maximum efficiency accurately; see Figure \ref{fig-RelativeError}.

\section{One-shot approximation $Z_\gen^{(0)}$, $\tau_{\rm lin}^{(0)}$ and $\beta_{\rm lin}^{(0)}$}\label{sec-one-shot}
The computation of $Z_\gen$, $\tau$ and $\beta$ requires the exact temperature distribution.
But they can be estimated directly from the material properties. In this section we derive an approximate formula for $Z_\gen$, $\tau$ and $\beta$.
The idea is to use the temperature distribution for $J=0$, which is similar to the exact temperature distribution because most devices induce small $J$ due to the small $zT$. Let $T^{(0)}$ be the temperature distribution for $J=0$ and define
\begin{alignat*}{3}
\rhoBar^{(0)} &:= \frac{1}{L} \int_{0}^{L} \rho(T^{(0)}(x)) \,dx &=& \frac{A}{L} R^{(0)}, \\
\frac{1}{\kappaBar^{(0)}} &:= \frac{1}{L} \int_{0}^{L} \frac{1}{\kappa (T^{(0)}(x))} \,dx &=& \frac{A}{L} \frac{1}{K^{(0)}}.
\end{alignat*}
From \eqref{SI-TEQ-1D} with $J=0$, we can check that
\begin{equation} \label{SI-heat-flux-approx}
-\kappa(T^{(0)}(x)) \frac{d T^{(0)}}{dx}(x) = \kappaBar^{(0)} \frac{\Delta T}{L}.
\end{equation}
Hence
\[
\begin{split}
\int_{T_c}^{T_h} \rho(T) \kappa(T)\,dT &= \int_{T_c}^{T_h} \rho(T^{(0)}) \Big(-\frac{\Delta T}{L} \kappaBar^{(0)}\Big) \frac{dx}{dT^{(0)}} \,dT^{(0)} \\
&= \frac{\Delta T}{L} \int_0^L \rho(T^{(0)}(x)) \,\kappaBar^{(0)} \,dx \\
&= \Delta T \,\rhoBar^{(0)} \,\kappaBar^{(0)}.
\end{split}
\]
Replacing $T$ with $T^{(0)}$ in $Z_\gen = \frac{\alphaBar^2}{\rhoBar\,\kappaBar}$, we have an one-shot approximation for $Z_\gen$:
\begin{equation}\label{one-shot-Zgen}
Z_\gen \approx \frac{\alphaBar^2}{\rhoBar^{(0)}\,\kappaBar^{(0)}} = \frac{ \left( \int \alpha \,dT \right)^2 }{\Delta T \,\int \rho \kappa \,dT} =: Z_\gen^{(0)}.
\end{equation}

\emph{To approximate $\tau$, we assume} the Seebeck coefficient is a linear function of $T$:
\[ \alpha(T) \approx \alpha_\lin(T) := \alpha_h + \left( \frac{\alpha_c - \alpha_h}{T_c - T_h} \right) \left(T- T_h \right). \]
In this way we can observe the effect of the gradient of $\alpha$ on $\tau$ more clearly. Since the $\tau$ in \eqref{SI-tau} has $K\,\delta T^{(1)}$ term, we estimate a relevant term:
\[
\begin{split}
F_T^{(1)}(s) &\approx \int_0^s \frac{1}{A} T \frac{d\alpha_\lin}{dT}(T(x)) \frac{dT}{dx}\,dx = \int_{T_h}^{T(s)} \frac{1}{A} T \frac{\alpha_c-\alpha_h}{T_c-T_h} \,dT\\
&= \frac{1}{2A} \frac{\alpha_c-\alpha_h}{T_c-T_h} (T(s)^2 - T_h^2) =: \widehat{F^{(1)}}(T(s)).
\end{split}
\]
\emph{Using} $-\kappa \frac{dT}{dx} \approx \kappaBar^{(0)} \frac{\Delta T}{L}$ from \eqref{SI-heat-flux-approx},
\[
\begin{split}
\delta T^{(1)} &= \int_0^L \frac{F_T^{(1)}(x)}{\kappa(x)}\,dx \approx - \int_0^L \frac{\widehat{F^{(1)}}(T(x))}{\kappaBar^{(0)}} \frac{L}{\Delta T}  \frac{dT}{dx}\,dx\\
&= \frac{1}{\kappaBar^{(0)}} \frac{L}{\Delta T} \int_{T_c}^{T_h} \widehat{F^{(1)}}(T)\,dT \\
&= \frac{1}{2 K^{(0)}} \frac{1}{\Delta T} \frac{\alpha_c-\alpha_h}{T_c-T_h} \frac{1}{3} (\Delta T)^2 (-3T_h +\Delta T) \\
&= \frac{\alpha_h-\alpha_c}{6 K^{(0)}} (-3T_h +\Delta T)=: \widehat{\delta T^{(1)}}
\end{split}
\]
where $K^{(0)} := \frac{A}{L}\kappaBar^{(0)}$. Therefore we have an one-shot approximation for $\tau$:
\[
\begin{split}
\tau &\approx \frac{1}{\overline{\alpha_\lin} \Delta T} \left[ (\overline{\alpha_\lin} - \alpha_h) T_h -K^{(0)} \,\widehat{\delta T^{(1)}} \right]\\
&= -\frac{1}{3} \frac{\alpha_h-\alpha_c}{\alpha_h+\alpha_c} =: \tau_{\rm lin}^{(0)}.
\end{split}
\]

\emph{To approximate $\beta$, we assume} the $\rho \kappa$ is a linear function of $T$:
\[ (\rho\kappa)(T) \approx (\rho\kappa)_\lin(T) := (\rho\kappa)_h + \left( \frac{(\rho\kappa)_c - (\rho\kappa)_h}{T_c - T_h} \right) \left(T- T_h \right). \]
\emph{Using} $-\kappa \frac{dT}{dx} \approx \kappaBar^{(0)} \frac{\Delta T}{L}$ from \eqref{SI-heat-flux-approx}, we approximate relevant terms for $\beta$:
\[
\begin{split}
F_T^{(2)}(s) &= \int_0^s \frac{1}{A^2}(\rho\kappa)(T(x)) \frac{1}{\kappa(x)} \,dx \approx \frac{-L}{A^2 \kappaBar^{(0)} \Delta T} \int_0^s (\rho\kappa)_\lin(T(x)) \frac{dT}{dx}\,dx\\
&= \frac{-L}{A^2 \kappaBar^{(0)} \Delta T} \int_{T_h}^{T(s)} (\rho\kappa)_\lin(T) \,dT \\
&= \frac{-L}{A^2 \kappaBar^{(0)} \Delta T} \Big[ (\rho\kappa)_h (T(s)-T_h) + \frac{1}{2} \frac{(\rho\kappa)_c-(\rho\kappa)_h}{T_c-T_h} (T(s)-T_h)^2 \Big]\\
& =: \widehat{F^{(2)}}(T(s))
\end{split}
\]
hence
\[
\begin{split}
\delta T^{(2)} &= \int_0^L \frac{F_T^{(2)}(x)}{\kappa(x)}\,dx \approx \int_0^L \widehat{F^{(2)}}(T(x)) \Big(-\frac{L}{\kappaBar^{(0)} \Delta T} \Big) \frac{dT}{dx}\,dx\\
&= \frac{-L}{\kappaBar^{(0)} \Delta T} \int_{T_h}^{T_c} \widehat{F^{(2)}}(T)\,dT \\
&= \frac{1}{6(K^{(0)})^2} \big( 2(\rho\kappa)_h + (\rho\kappa)_c \big) =: \widehat{\delta T^{(2)}}.
\end{split}
\]
Therefore we have an one-shot approximation for $\beta$:
\[
\begin{split}
\beta &\approx \frac{2}{\frac{L}{A}\rhoBar^{(0)}} K^{(0)} \,\widehat{\delta T^{(2)}} - 1 = \frac{1}{3 \,\rhoBar^{(0)}\kappaBar^{(0)}} (2(\rho\kappa)_h +(\rho\kappa)_c) -1\\
&\approx \frac{1}{\frac{3}{2} ((\rho\kappa)_h+(\rho\kappa)_c)} (2(\rho\kappa)_h +(\rho\kappa)_c) -1\\
&= \frac{1}{3} \frac{(\rho\kappa)_h - (\rho\kappa)_c}{(\rho\kappa)_h +(\rho\kappa)_c} =: \beta_{\rm lin}^{(0)}.
\end{split}
\]
In summary, we have one-shot approximations as following:
\begin{equation}\label{one-shot-approx}
Z_\gen \approx Z_\gen^{(0)} \equiv \frac{ \left( \int \alpha \,dT \right)^2 }{\Delta T \,\int \rho \kappa \,dT}, 
\quad
\tau \approx \tau_{\rm lin}^{(0)} \equiv -\frac{1}{3}\frac{\alpha_h - \alpha_c}{\alpha_h+\alpha_c}, \quad 
\beta \approx \beta_{\rm lin}^{(0)} \equiv \frac{1}{3}\frac{\rho_h\kappa_h-\rho_c\kappa_c}{\rho_h\kappa_h+\rho_c\kappa_c}.
\end{equation}

\begin{figure}
\centering \includegraphics[width=0.8\textwidth]{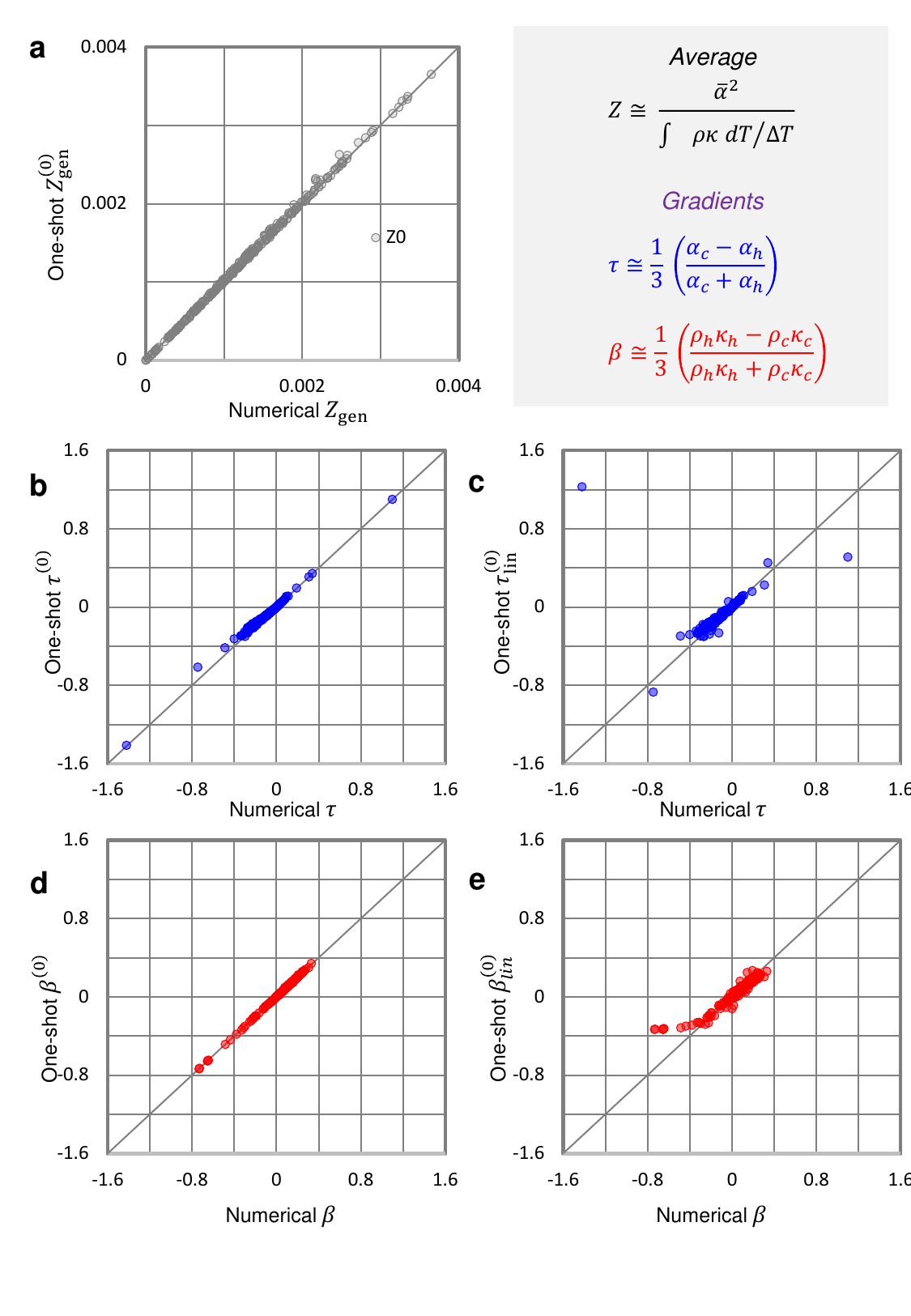}
\caption{Estimation of thermoelectric degrees of freedom for 276 materials. Numerical $Z_\gen$, $\tau$, $\beta$ are computed using the exact $T$ at the maximum efficiency. One-shot approximations $Z_\gen^{(0)}$, $\tau^{(0)}$, $\beta^{(0)}$ are computed using the $T^{(0)}$ for $J=0$. Going further, the $\tau_{\rm lin}^{(0)}$ and $\beta_{\rm lin}^{(0)}$ are computed by assuming the linearity of $\alpha$ and $\rho\kappa$; see \eqref{one-shot-approx} for their explicit formula.
}\label{fig-one-shot-approx}
\end{figure}

The \emph{one-shot approximation} derived above is accurate enough for many cases.
See Figure \ref{fig-one-shot-approx}, where we compare the exact $Z_{\rm gen}$, $\tau$, $\beta$ with their one-shot approximations for 276 materials.

\begin{figure}
\centering \includegraphics[width=\textwidth]{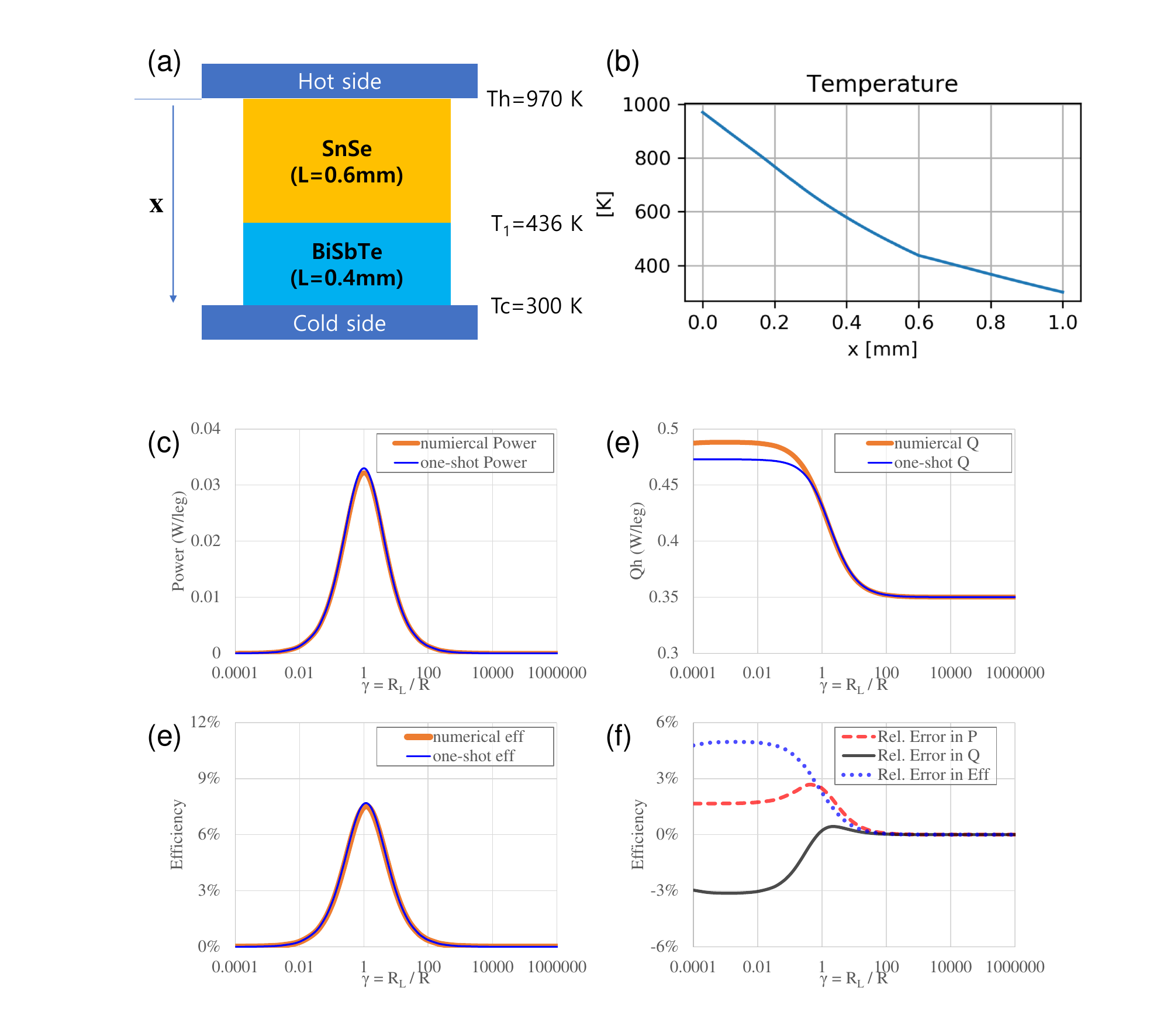}
\caption{
The thermoelectric performances of a two-stage segmented leg predicted by the one-shot approximation. The numerical exact values are computed by fixed-point iteration and the one-shot values are computed using $Z_\gen^{(0)}$, $\tau_{\rm lin}^{(0)}$ and $\beta_{\rm lin}^{(0)}$; see \eqref{one-shot-approx} for the explicit one-shot formula.
(a) The geometry of the segmented leg: $\rm SnSe$ \cite{zhao_ultralow_2014} and $\rm BiSbTe$ \cite{poudel_high-thermoelectric_2008} are used for hot and cold-side materials. $T_h = 970 K$ and $T_c = 300 K$ are used.
(b) Exact temperature distribution obtained by solving the integral equation \eqref{SI-T-integral-form} of $T$ with fixed-point iteration.
(c) Power delivered outside, (d) heat current at the hot side, (e) efficiency, and (f) relative errors in power, heat current, efficiency between the numerical value and the one-shot approximation.
}\label{fig-segQ}
\end{figure}

Furthermore, these one-shot approximations can be used to predict the performance of \emph{segmented} devices.
In Figure \ref{fig-segQ}, we consider a two-stage segmented leg with no contact resistance. The segmented leg consists of SnSe \cite{zhao_ultralow_2014} for hot side and BiSbTe \cite{poudel_high-thermoelectric_2008} for cold side.
The exact temperature distribution $T$ insdie the leg shows a jump of the gradient at $x=0.6$ due to the inhomogeneity of the material; see Figure \ref{fig-segQ}(b).
Despite the nonlinearity of the $T$, the one-shot approximation using $Z_\gen^{(0)}$, $\tau_{\rm lin}^{(0)}$ and $\beta_{\rm lin}^{(0)}$, which does not use the exact $T$, shows high accuracy in prediction of thermoelectric performances; see Figure \ref{fig-segQ}(c)-(f).
The relative error is high near $\gamma=0$, where the reaction term is large due to the large electric current. For large $\gamma$, the error is negligible. Near the $\gamma=1$, the error is acceptable; the relative error is less than 5\%. The one-shot approximation predicts the maximum efficiency to be 7.68\% while the exact value is 7.53\%.

\clearpage

\section{Maximum efficiency prediction using $\eta_{\rm max}^{\rm gen}$ }

\begin{figure}[h]
\centering \includegraphics[width=\textwidth]{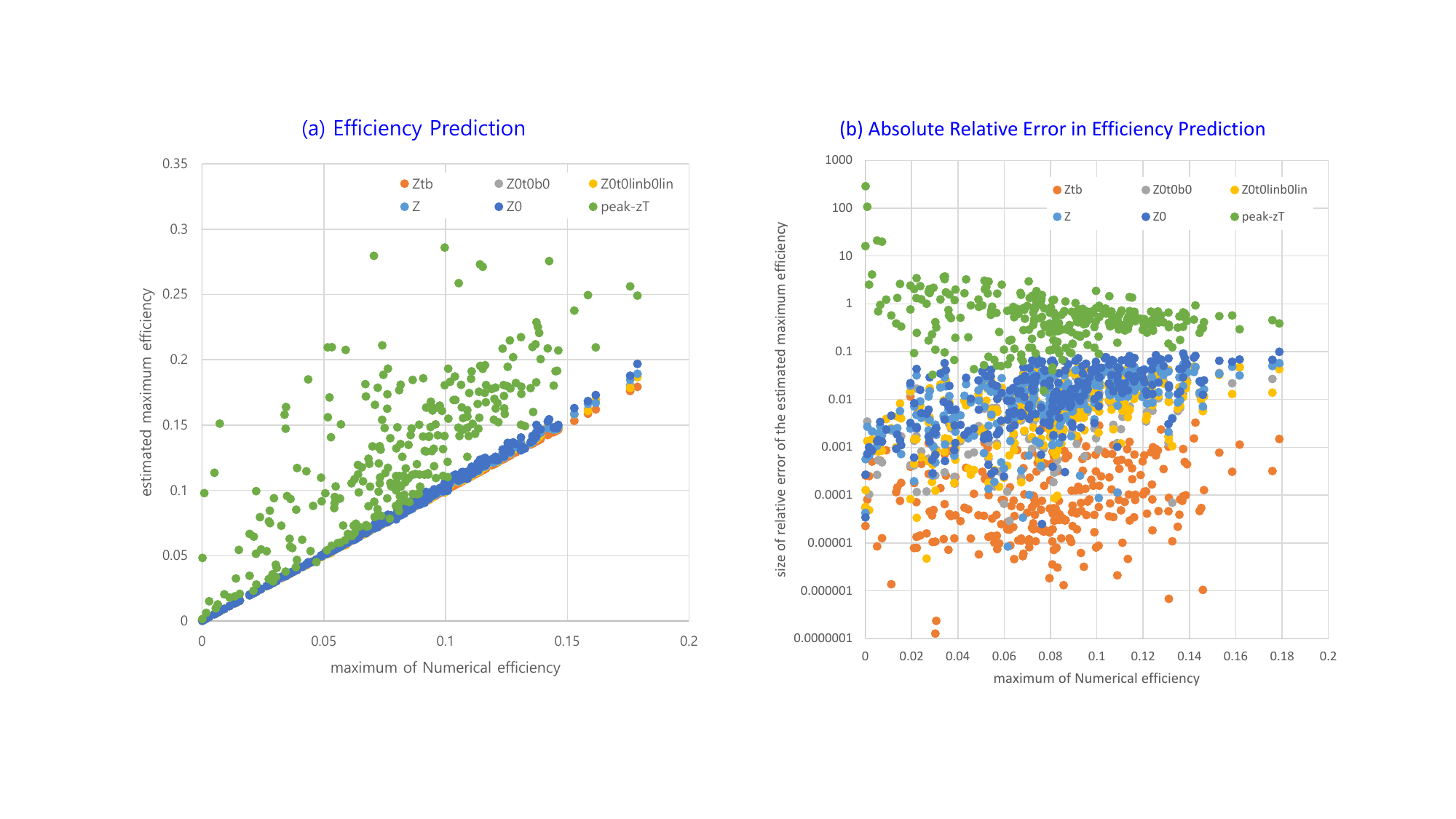}
\caption{Efficiency estimation for 276 materials using the formula $\eta_{\rm max}^{\rm gen} (Z_\gen, \tau, \beta)$ in \eqref{SI-max-efficiency} and the peak $zT$. In use of $\eta_{\rm max}^{\rm gen} (Z_\gen, \tau, \beta)$, there are five options: using (i) exact $Z_\gen, \tau, \beta$ (Ztb), (ii) $Z_\gen^{(0)}, \tau^{(0)}, \beta^{(0)}$ (Z0t0b0), (iii) $Z_\gen^{(0)}, \tau_{\rm lin}^{(0)}, \beta_{\rm lin}^{(0)}$ (Z0t0linb0lin), (iv) exact $Z_\gen$ only, $\tau=0$, $\beta=0$ (Z), (v) $Z_\gen^{(0)}$ only, $\tau=0$, $\beta=0$ (Z0).
(a) Comparison of the exact maximum efficiency and the estimations. (b) Relative error of the estimations (absolute error divided by the exact maximum efficiency). The peak $zT$ is malfunctioning in efficiency prediction; the relative error can be over $100\%$. On the other hand, the formula $\eta_{\rm max}^{\rm gen} (Z_\gen, \tau, \beta)$ has small relative error; even the simplest formula with $Z_\gen^{(0)}, \tau_{\rm lin}^{(0)}, \beta_{\rm lin}^{(0)}$ has the standard error (=root mean square of relative errors) less than 2\%. See Table \ref{Table-rel-err-eff-formula} for detailed values.
}\label{fig-RelativeError}
\end{figure}

In Figure \ref{fig-RelativeError}, we can observe that the maximum efficiency estimation formula $\eta_{\rm max}^{\rm gen} (Z_\gen, \tau, \beta)$ in \eqref{SI-max-efficiency} is highly accurate.
In Table \ref{Table-rel-err-eff-formula}, various statistics on the relative error of maximum efficiency ($\frac{\eta_{\rm max}^{\rm gen}  -  \eta_{\rm max}}{ \eta_{\rm max}  }$) are given.
\begin{table}[h]
\caption{Statistics on the relative error (RelErr) of the maximum efficiency estimation formula $\eta_{\rm max}^{\rm gen} (Z_\gen, \tau, \beta)$ in \eqref{SI-max-efficiency}. Average (Avg), root mean square (RMS RelErr or StdErr), maximum (max), and minimum (min) of the relative errors are estimated for 276 materials for thermoelectric power generator working at their available temperature.} \label{Table-rel-err-eff-formula}
\begin{tabular}{|c|r|r|r|r|r|r|}
\hline
\multirow{3}{*}{\begin{tabular}[c]{@{}c@{}}276 materials\\ for power\\ module\end{tabular}} & \multicolumn{6}{c|}{Relative error in maximum efficiency formula} \\ \cline{2-7} 
 & \multicolumn{5}{c|}{$\eta_{\rm max}^{\rm gen}$} & \multicolumn{1}{c|}{$\eta_{\rm max}^{\rm const}$} \\ \cline{2-7} 
 & \multicolumn{1}{c|}{$Z_\gen,\tau,\beta$} & \multicolumn{1}{c|}{$Z_\gen^{(0)},\tau^{(0)},\beta^{(0)}$} & \multicolumn{1}{c|}{$Z_\gen^{(0)},\tau_{\rm lin}^{(0)},\beta_{\rm lin}^{(0)}$} & \multicolumn{1}{c|}{$Z_\gen$} & \multicolumn{1}{c|}{$Z_\gen^{(0)}$} & \multicolumn{1}{c|}{peak $zT$} \\ \hline
Avg RelErr & 0.02\% & 1.11\% & 1.08\% & 1.42\% & 2.29\% & 235\% \\ \hline
StdErr (RMS RelErr) & 0.09\% & 1.38\% & 1.38\% & 1.52\% & 2.47\% & 1854\% \\ \hline
max RelErr & 1.15\% & 5.45\% & 5.23\% & 5.80\% & 9.96\% & 28835\% \\ \hline
min RelErr & -0.61\% & -1.92\% & -1.76\% & -1.78\% & -2.48\% & -4\% \\ \hline
\end{tabular}
\end{table}

If we use the exact $Z_\gen, \tau, \beta$, the standard error (=root mean square of relative errors) of $\eta_{\rm max}^{\rm gen}$ is $9.60 \times 10^{-4}$.

If we use $Z_\gen^{(0)}, \tau_{\rm lin}^{(0)}, \beta_{\rm lin}^{(0)}$, the standard error is $1.75 \times 10^{-2}$. 
For the signle crystalline $\rm SnSe$ with peak $zT$ of 2.6, the relative error of one shot method is found to be only $6.82 \times 10^{-3}$.
However, when we use the different approximation such as linear $T(x)$ or different average scheme for $z$, the error becomes larger than ours due to the non-linearity of $T$ for this material \cite{kim2015relationship}.

If we only use the $Z_\gen^{(0)}$ with zero $\tau$ and $\beta$, 
the efficiency is still well predicted with the standard error of  $3.37 \times 10^{-2}$.
But, in some materials, the error is relatively large due to the neglect of the $\tau$ and $\beta$. The largest relative error of 10\% is found for \cite{wu_broad_2014}, due to the non-vanishing gradient parameters 
($\tau = -0.222 \approx \tau^{(0)} = -0.177 \approx  \tau_{\rm lin}^{(0)} = -0.204$,
$\beta = 0.2085 \approx \beta^{(0)} = 0.228 \approx  \beta_{\rm lin}^{(0)} = 0.185$, when $T_h = 918 K$ and $T_c = 304 K$).

\section{Efficiency rank estimation using $Z_\gen^{(0)}$} \label{sec-eff-rank}

The $Z_\gen$ is a figure of merit, so the bigger $Z_\gen$ usually implies the bigger maximum efficiency.
Then if we rank TE devices in order of $Z_\gen$, will we get the correct rank in order of exact maximum efficiency?
To measure such an effect quantitatively, we define the \emph{top-rank-preserving probability} by the ratio of the number of correct top ranks predicted by some estimation parameter, to the total number of top ranks.

In Table \ref{Table-rank-preserving}, we observe the top-rank-preserving probability is high even if we use the simplest estimation $Z_\gen^{(0)}$.
We computed the maximum thermoelectric efficiency of 5-stage segmented leg for all possible configuration using 18 candidates materials in Table \ref{table-18-mats-temp-range}. Thus there are $18^5 = 1,889,568$ device structures.
No contact resistance is imposed, but it can be easily imposed in our numerical schme by adding a stage with zero Seebeck coefficient.
The result shows with the 82\% probability, the top 1\% rank configurations in order of exact maximum efficiency can be found in the top 1\% ranks in order of $Z_\gen^{(0)}$. Hence one may perform faster high-throughput screening by computing $Z_\gen^{(0)}$ only, without having to compute the numerical maximum efficiency.

\begin{table}[h]
\caption{Comparison of top ranks in order of exact maximum efficiency and estimation parameters. The Top-rank-preserving probability means the ratio of the number of correct top ranks predicted by the estimation parameter, to the total number of top ranks.
The 18 candidates materials in Table \ref{table-18-mats-temp-range} are used to generate 5-stage segmented legs. Each stage of the leg has the same cross sectional area $1 \,{\rm mm^2}$ and the same length $1/5 \,{\rm mm}$ (total length is $1 \,{\rm mm}$). The hot- and cold-side temperatures are $T_h=900K$ and $T_c=300K$.
%\textcolor{red}{JC: please fill the $Z_\gen$ category.}
} \label{Table-rank-preserving}
\begin{tabular}{|c|c|c|c|c|}
\hline
\multicolumn{2}{|c|}{\multirow{2}{*}{Rank}} & \multicolumn{3}{c|}{Top-Rank-Preserving Probability} \\ \cline{3-5} 
\multicolumn{2}{|c|}{}                      & \multicolumn{1}{c|}{$Z_\gen$} & \multicolumn{1}{c|}{$\eta_{\rm max}^{\rm gen}(Z_\gen^{(0)}, \tau^{(0)}, \beta^{(0)})$} & \multicolumn{1}{c|}{$Z_\gen^{(0)}$}  \\ \hline
Top 0.1\%            & $<$1,891      &   87\%           &    73\%      & 73\%       \\ \hline
Top 1\%               & $<$18,897    &   90\%           &    84\%      & 82\%       \\ \hline
Top 2\%               & $<$37,792    &   93\%           &    88\%      & 86\%       \\ \hline
Top 4\%               & $<$75,584    &   94\%            &   89\%      & 90\%       \\ \hline
All configurations & 1,889,568     & 100\%           &  100\%    & 100\%     \\ \hline
\end{tabular}
\end{table}

\subsubsection*{Additional information}
The best efficiency in the setting of Table \ref{Table-rank-preserving} is 21.95\% while the one-shot approximation $\eta_{\rm max}^{\rm gen}(Z_\gen^{(0)}, \tau^{(0)}, \beta^{(0)})$ predicts it would be 22.30\%. For top 100,000 configurations, the root mean square error is 0.0415.

\subsubsection*{Computation algorithm}
The maximum thermoelectric conversion efficiency of a given device configuration is computed using the following procedures.
\begin{enumerate}
\item Prepare the thermoelectric property curves using digitized data. Each curve is linearly interpolated at intermediate temperature and extrapolated as constant values at the end point temperatures.
\item Choose the linear function as the initial guess $T_0$ of exact temperature distribution.
\item \label{step-repeat} Given a temperature distribution $T_n$, compute thermoelectric degrees of freedom using the definition in \eqref{def-Zgen} and \eqref{SI-tau}. Then estimate the optimal current density $J$ using the formula \eqref{gamma-gen-max}. If a given structure is segmented, the material properties are position-dependent as well as temperature-dependent (but there is no additional difficulty in computation).
\item Compute $T_{n+1}$ by evaluating the right-hand side of the integral equation \eqref{SI-T-integral-form}.
\item If $T_{n+1}$ agrees with $T_n$, go to the next step. Otherwise replace $T_n$ by $T_{n+1}$ and go back to the step \ref{step-repeat}.
\item Using the converged temperature distribution $T_{n+1}$, compute the maximum efficiency from $\eta_{\rm max}^{\rm gen}(Z_\gen, \tau, \beta)$.
\end{enumerate}

\subsubsection*{Computation time} In a single core computer, the computation of the maximum efficiency of a segmented leg takes less than 1 second. Thus, for total computation, it may take about 525 hours (22 days). We used a high-performance-computing (HPC) system consisting of 500 processors so the computation took about 1 hour.

\section{Why peak $zT$ fails for $\rm BiSbTe$-like and $\rm SnSe$-like materials}

While the peak $zT$ of $\rm SnSe$-like materials is significantly greater than that of $\rm BiSbTe$-like materials, the efficiency of the latter is significantly greater than the former ($\rm SnSe$ has the highest peak $zT$ of 2.6 at 923 K); see Figure 1 in the paper. This extreme failure case of $zT$ can be explained by our additional figure of merit $\tau$.

\begin{figure}[h]
\centering\includegraphics[width=\textwidth]{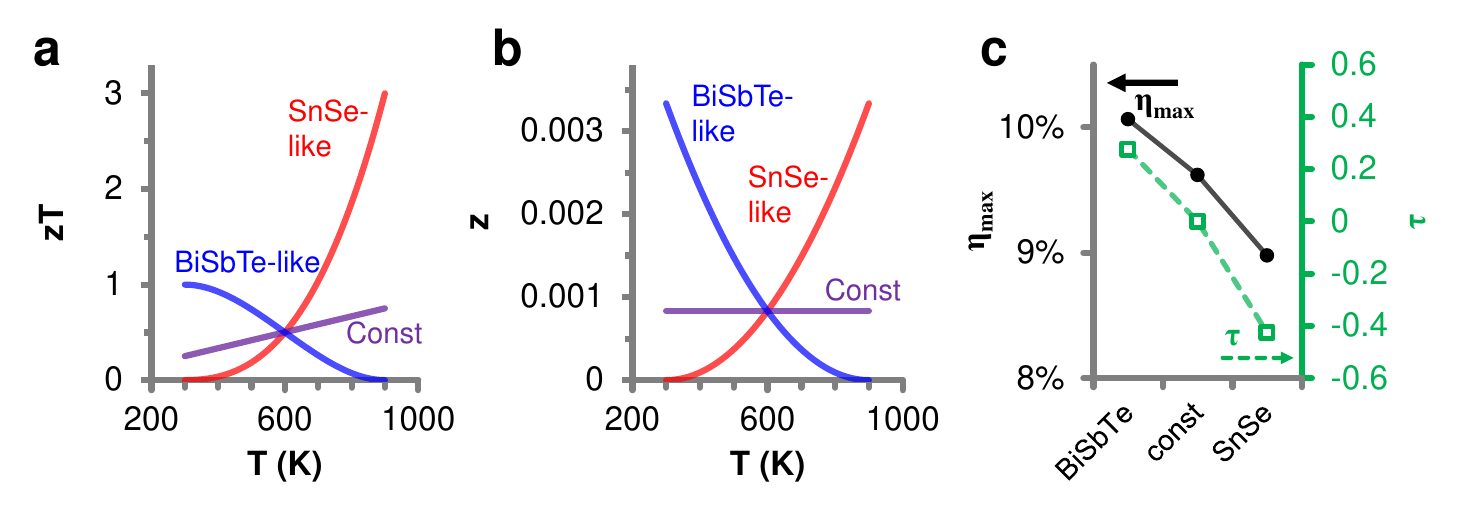}
\caption{The $zT$, the maximum efficiency, and $\tau$ for three imaginary materials which imitates $\rm BiSbTe$-like, $\rm SnSe$-like, and constant-$z$ materials. The $\alpha$ of the materials is linear while the $\rho$ and $\kappa$ of them are constant. The materials have the same $\alphaBar$ and $Z_\gen$. For working temperature from $300 K$ to $900 K$, the highest maximum efficiency is found in the $\rm BiSbTe$-like material due to the positive $\tau$.
%\textcolor{red}{JC: remove `beyondZT', remove unnecessary blanks, clarify the symbol for $\eta_{\rm max}$ and $\tau$ in Figure (c), using color or label.}
}\label{fig-eg-zT-fail}
\end{figure}

Consider three imaginary materials imitaing $\rm BiSbTe$-like, $\rm SnSe$-like, and constant-$z$ materials. For simplicity, we impose some assumptions on their material properties. The $\rho$ and $\kappa$ of them are temperature-independent and they have the same $\alphaBar$. The $\alpha$ of them is linear on temperature; the $\rm BiSbTe$-like material has linearly decreasing $\alpha$, the $\rm SnSe$-like material has linearly increasing $\alpha$, and the constant-$z$ material has the constant $\alpha$. Then, as shown in Figure \ref{fig-eg-zT-fail}, the peak $zT$ of the $\rm SnSe$-like material is very high. However, due to the temperature-dependent profile of $\alpha$, the $\tau$ of the $\rm SnSe$-like material is negative while the $\tau$ of $\rm BiSbTe$-like material is positive; see \eqref{one-shot-approx}. Since the $Z_\gen$ is the same for the three materials, the $\tau$ is the main figure of merit which concludes that the $\rm BiSbTe$-like material has higher maximum efficiency than the $\rm SnSe$-like material. This example shows the gradient of material properties can affect the maximum efficiency.

\section{Optimal doping concentration for $\rm Bi_2Te_3$}

In this section, using \emph{calculated} material properties, we design funtionally graded materials (FGM) composed of $\rm Bi_2 Te_3$ to maximize the efficiency. The thermoelectric properties are calculated using the density functional theory (DFT) \cite{hohenberg1964inhomogeneous,kohn1965self} combined with the Boltzmann transport equation.
For the DFT calculations, we use the generalized gradient approximation (GGA) parameterized by PBE(Perdew, Burke, and Ernzerhof) \cite{perdew1996generalized}, and the projector augmented-wave (PAW) pseudopotential \cite{blochl1994projector}; both of them are implemented in the VASP code \cite{kresse1996g,kresse1999ultrasoft}. The experimental lattice parameters  for $\rm Bi_2Te_3$ are used, while the internal coordinates are fully relaxed. The electronic band structure is calculated using the spin-orbit interaction. The $k$-point mesh of $36 \times 36 \times 36$ is used.
The electronic transport properties are predicted using the DFT band structure coupled with the Boltzmann transport equation within a rigid band approximation and the constant relaxation time approximation; they are implemented in BoltzTraP code \cite{madsen2006boltztrap,ryu2017thermoelectric}. Note that we use the experimental band gap of 0.18 eV.
The phonon thermal conductivity is calculated using phono3py code \cite{phono3py,ryu2016computational}. 
The force constants are obtained from the 240-atom supercell with the two-atom displacements using VASP code with the single $k$-point $\Gamma$ and then the thrid-order phonon Hamiltonian is constructured.
The three phonon scattering rates are calculated using the Fermi's golden rule. We also include the effective boundary scattering of 10 nm in addition to the three-phonon scattering. 
Then the thermal conductivity is calculated by integrating the conductivity on the phonon $q$-point mesh of $11 \times 11 \times 11$.

\begin{center}
\begin{figure}[h]
\includegraphics[width=\textwidth]{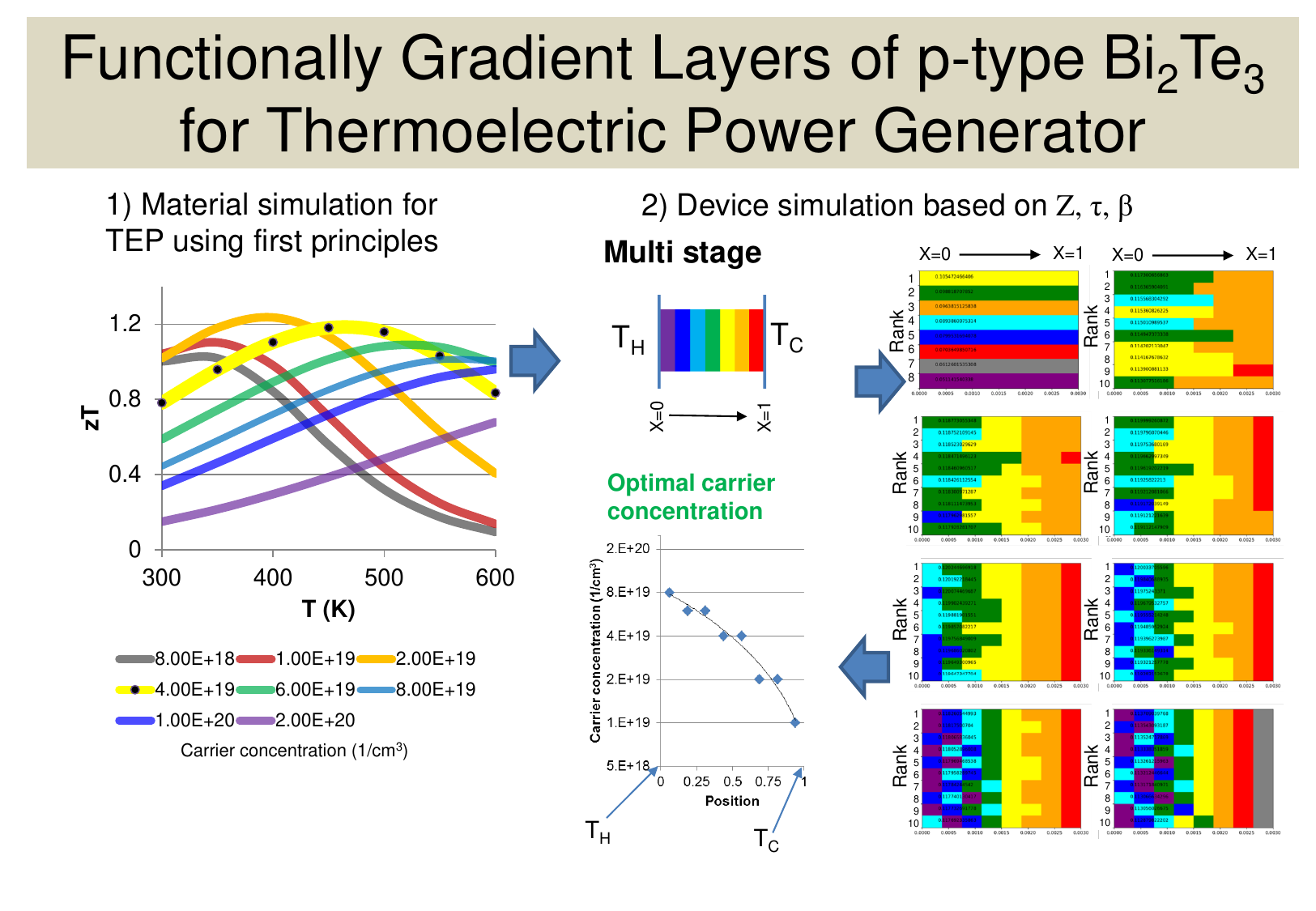}
\caption{Design process of functionally graded materials for thermoelectric power generator and its result. The temperature range from 300 K to 600 K is considered. High-throughput computation of efficiency is performed to search the optimal carrier doping concentration. (Left) Thermoelectric properties of $\rm Bi_2Te_3$ calculated by DFT. (Middle top) Schematic structure of segmented devices; different color means different doping concentration. (Right) Top 10 segmented structures when the number of stage (number of segmentation of equal length) is fixed; 1 to 8 stages are considered. (Middle bottom) Optimal carrier doping concentration having the highest efficiency.
}\label{fig-gradation1}
\end{figure}
\end{center}

\begin{center}
\begin{figure}[h]
\includegraphics[width=\textwidth]{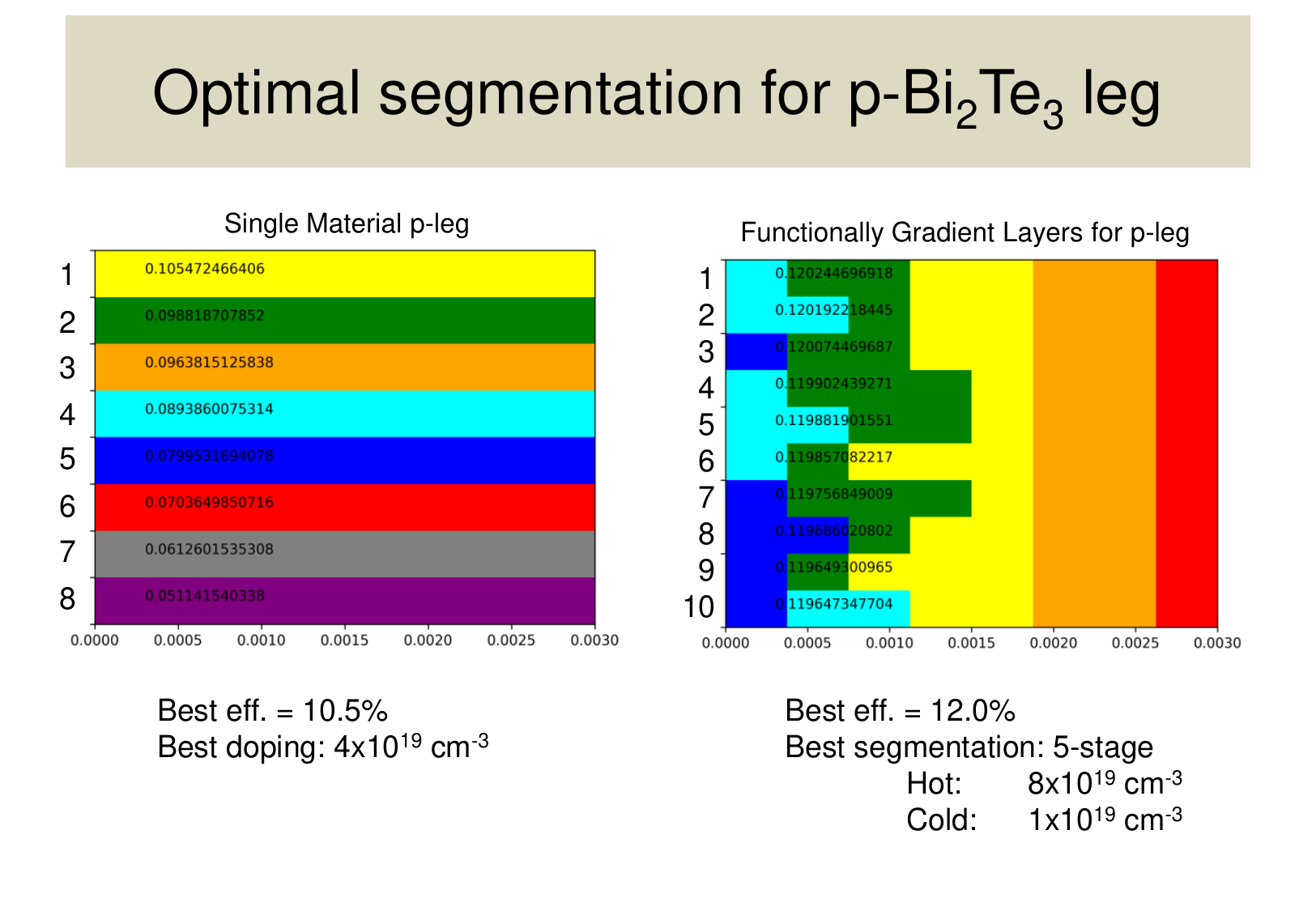}
\caption{Top 10 segmented structures ranked in order of maximum efficiency. (Left) No segmentation. (Right) 5-stage segmentation. Each color represents a distinct material. The top rank structure is shown in the first row: yellow is the first rank for no segmentation. The 1cyan-2green-2yellow-2orange-1red segmented structure in the first row of the right figure is optimal among the $8^8$ configurations with the highest efficiency of 12.0\%.
}\label{fig-gradation2}
\end{figure}
\end{center}

We calculate the maximum efficiency of functional gradient layers (FGL) based on $\rm Bi_2Te_3$ for temperature range from 300 K to 600 K. We consider various segmented devices having 1 stage to 8 stages with eight different carrier concentrations ($8 \times 10^{18}, ~1 \times 10^{19}, ~2 \times 10^{19}, ~4 \times 10^{19}, ~8 \times 10^{19}, ~1 \times 10^{20}, ~2 \times 10^{20} ~{\rm cm^{-3}}$).
We perform high-throughput computation to find the optimal segmented FGL.
There are $8^8$ possible configurations in total.
The temperature distribution inside a device is obtained by using fixed-point iteration of the integral equation \eqref{SI-T-integral-form}. At the same time, the current density is optimized to find the maximum efficiency; see \emph{Computation algorithm} in \S\ref{sec-eff-rank} for more details.
Figure \ref{fig-gradation1} shows the thermoelectric properties calculated by DFT, various segmented structures with its efficiency, and the optimal carrier concentration as a function of position.
Figure \ref{fig-gradation2} shows the highest efficiency is obtained for a 5-stage segmented device. For single stage, the maximum efficiency of 10.5 \% is found at the doping concentration $4 \times 10^{19} ~{\rm cm^{-3}}$. For multi-stage, the maximum efficiency is found at the 5-stage with the optimal carrier concentration varying from $8 \times 10^{19} \,{\rm cm^{-3}}$ to  $1 \times 10^{19} \,{\rm cm^{-3}}$ as going from hot to cold side.

\section*{Acknowledgement}
This work was supproted by Korea Electrotechnology Research Institute (KERI) Primary research program through the National Research Council of Science and Technology (NST) funded by the Ministry of Science and	 ICT (MSIT) of the Republic of Korea [No. 18-12-N0101-34 (Development of Design Tools of Thermoelectric and Energy Materials)]. This work was also supported by the Korea Institute of Energy Technology Evaluation and Planning (KETEP) and the Ministry of Trade, Industry and Energy (MOTIE) of the Republic of Korea [No. 20162000000910 (Development of High Performance Thermoelectric Modules by Power Modulation) and No. 20172010000830 (Developments of Thermoelectric Power Generation System using Unused Heats in Industry and Business Model)].

%\begin{thebibliography}{1}
%\bibitem{dummy} Articles are restricted to 50 references, Letters to 30.
%\bibitem{dummyb} No compound references -- only one source per reference.
%\end{thebibliography}

\bibliographystyle{naturemag}
\bibliography{TEPdata3_DFT}
%\printbibliography[type=article]